\documentclass[a4paper]{article}

\usepackage[english]{babel}
\usepackage[utf8]{inputenc}
\usepackage{amsmath}
\usepackage{amsthm}
\usepackage{amssymb}
\usepackage[semicolon]{natbib}
\usepackage{graphicx}
\usepackage{enumitem}
\usepackage{setspace}
\usepackage[colorinlistoftodos]{todonotes}
\usepackage[margin=1in]{geometry}
\usepackage{listings}
\usepackage[T1]{fontenc}
\usepackage{babel}
\usepackage{geometry}
\usepackage{titling}
\usepackage{blindtext}
\usepackage[draft]{hyperref}
\usepackage{tikz}
\usetikzlibrary{positioning}
\usepackage{caption}
\usepackage{float}
\usepackage{relsize}
\usepackage{subcaption}
\usepackage{bbm}
\usepackage[framemethod=tikz]{mdframed}
\usepackage{color}
\usepackage{ulem}

\tikzset{
  treenode/.style = {align=center, inner sep=0pt, text centered, font=\sffamily, minimum size=0.3cm},
  circnode/.style = {treenode, circle, white, font=\sffamily\bfseries, draw=black, fill=black}
}

\allowdisplaybreaks

\title{Calculating Higher-Order Moments of Phylogenetic Stochastic Mapping Summaries in Linear Time}
\author{
  Amrit Dhar$^1$ and Vladimir N. Minin$^{1,2}$ \\ \\
  $^1$Department of Statistics, University of Washington, Seattle \\
  $^2$Department of Biology, University of Washington, Seattle
}

\date{\today}

\begin{document}
\maketitle

\begin{abstract}
Stochastic mapping is a simulation-based method for probabilistically mapping substitution histories onto phylogenies according to continuous-time Markov models of evolution.
This technique can be used to infer properties of the evolutionary process on the phylogeny and, unlike parsimony-based mapping, conditions on the observed data to randomly draw substitution mappings that do not necessarily require the minimum number of events on a tree.
Most stochastic mapping applications simulate substitution mappings only to estimate the mean and/or variance of two commonly used mapping summaries: the number of particular types of substitutions (labeled substitution counts) and the time spent in a particular group of states (labeled dwelling times) on the tree.
Fast, simulation-free algorithms for calculating the mean of stochastic mapping summaries exist.
Importantly, these algorithms scale linearly in the number of tips/leaves of the phylogenetic tree.
However, to our knowledge, no such algorithm exists for calculating higher-order moments of stochastic mapping summaries.
We present one such simulation-free dynamic programming algorithm that calculates prior and posterior mapping variances and scales linearly in the number of phylogeny tips.
Our procedure suggests a general framework that can be used to efficiently compute higher-order moments of stochastic mapping summaries without simulations.
We demonstrate the usefulness of our algorithm by extending previously developed statistical tests for rate variation across sites and for detecting evolutionarily conserved regions in genomic sequences.
\end{abstract}


\section{Introduction}

Given a multiple sequence alignment of DNA nucleotides, scientists are often interested in reconstructing a phylogenetic tree to help them learn more about the ancestral relationships between the sequences and the underlying evolutionary process \citep{yang2006computational}.
However, in some cases, phylogeny estimation by itself does not provide all the needed information about sequence evolution because we observe data only at the tips of the phylogeny.
We do not have much insight into the evolution taking place on the different tree branches other than through the estimated branch lengths, which are usually specified in terms of the expected number of substitutions per site \citep[Chapter 13]{felsenstein2004inferring}.
However, researchers are often interested in making inferences about the evolutionary process on the phylogeny because these inferences could be used to answer important scientific questions.
For instance, estimates of non-synonymous/synonymous substitution rate ratios on a phylogeny are commonly used to test for positive selection on protein-coding genes \citep{nielsen1998likelihood}.
Stochastic mapping can be used to accurately estimate these ratios and, more generally, can help us make reliable inferences about the latent evolutionary process on the phylogeny \citep{nielsen2002mapping, huelsenbeck2003stochastic, dimmic2005detecting, zhai2007exploring, lemey2012counting}.
Stochastic mapping is a simulation-based technique used to probabilistically map substitution histories onto phylogenies according to continuous-time Markov chain (CTMC) models of evolution.
This approach was motivated by the need for alternatives to parsimony mapping, which focuses attention on mappings requiring the fewest substitutions.
\par
Stochastic mapping was first introduced by \citet{nielsen2002mapping}, who described how to sample substitution mappings from the posterior probability distribution of mappings for a single trait or a site in a multiple sequence alignment.
By using Nielsen's sampling procedure, one can compute Monte Carlo estimates for the posterior mean and/or variance, among other properties of the posterior distribution, of any mutational mapping summary random variable of interest.
The two most popular stochastic mapping summaries are the number of particular types of substitutions (labeled substitution counts) and the time spent in a particular group of states (labeled dwelling times) on the tree.
In most applications, substitution mappings are simulated only to estimate the mean and/or variance of the two mapping summaries discussed above \citep{minin2008fast}.
Recognizing this, \citet{minin2008fast} synthesized previous work of \citet{hobolth2005statistical}, \citet{dutheil2005model}, and \citet{holmes2002expectation}, among others, and developed an efficient algorithm that analytically calculates the expectations of the aforementioned mapping summaries.
The authors compute restricted expectations of CTMC labeled substitution counts \citep{ball2005simple, minin2008counting} and labeled dwelling times \citep{neuts1995algorithmic, guindon2004modeling, minin2008fast} on each tree branch and propagate these expectations across the phylogeny using a generalized pruning algorithm \citep{felsenstein1981evolutionary}.
Similarly to Felsenstein's pruning algorithm, the algorithm of \citet{minin2008fast} scales linearly in the number of phylogeny tips.
\citet{minin2008fast} drew inspiration from the work of \citet{schadt1998computational}, who formulated a similar algorithm that computes first derivatives of phylogenetic likelihood functions.
Unfortunately, it is not straightforward to extend the algorithm of \citet{minin2008fast} to efficiently calculate the variances of the previously mentioned mapping summaries; as a result, these stochastic mapping variances can only be approximated using Monte Carlo simulations.
\par
In this paper, we present a simulation-free dynamic programming algorithm that calculates prior and posterior mapping variances and scales linearly in the number of phylogeny tips.
We draw upon concepts introduced by \citet{kenney2012hessian}, who developed a recursive procedure for calculating second derivatives of phylogenetic likelihood functions.
Our procedure suggests a general framework that can be used to efficiently compute higher-order moments of stochastic mapping summaries without simulations.
The structure of the rest of the paper is as follows.
Section 2 introduces notation that is used throughout the entire paper and discusses our research problem more formally.
In Section 3, we give a description of our algorithm for efficiently calculating these stochastic mapping variances.
In Section 4, we demonstrate the usefulness of our algorithm by extending previously developed statistical tests for rate variation across sites and for detecting evolutionarily conserved regions in genomic sequences.
Concluding remarks are presented in Section 5.

\section{Notation and Problem Background}

We use much of the notation provided by \citet{minin2008fast}.
Suppose we have a discrete evolutionary trait $X$ (i.e.\ DNA base) that takes on $m$ distinct states and that evolves according to a CTMC on a phylogeny.
This evolutionary process, call it $\psi_{\boldsymbol{\theta}}$, depends on the parameter set $\boldsymbol{\theta} = \{ \tau, \mathbf{t}, \mathbf{Q}, \boldsymbol{\pi} \}$, which consists of a rooted tree topology $\tau$ with $n$ tips and $B_n = 2n-2$ branches; branch lengths $\mathbf{t} = (t_1, ..., t_{B_n})$; a reversible CTMC rate matrix $\mathbf{Q} = \{ q_{ij} \}$ for $i,j = 1, ..., m$; and a CTMC stationary distribution $\boldsymbol{\pi} = (\pi_1, ..., \pi_m)^T$.
We assume that our evolutionary process starts at stationarity (i.e.\ we assume that the root distribution is equal to $\boldsymbol{\pi}$).
While not necessary, this commonly used assumption ensures that the stochastic mapping moments will be invariant to the placement of the root \citep{minin2008fast}.
When this assumption is not used, as often is the case in analyses of morphological traits \citep{pagel1999maximum}, our methods still work without modification, but the root of the tree has to be specified by the user.
Matrix $\mathbf{P}(t) = \{ p_{ij}(t) \} = \exp(\mathbf{Q}t)$ represents the CTMC transition probability matrix for a branch of length $t$.
\par
We define $\Theta = \{ 1, ..., B_n \}$ to be the set of branch indices of $\tau$.
Let $\Theta_b = \{ b^* \in \Theta \ | \ b^* \preceq b \}$ denote the set of branch indices in the subtree relating all descendants of branch $b$, including $b$, where $b^* \preceq b$ for $b,b^* \in \Theta$ if either $b^*$ is a descendant of $b$ or $b^*$ is equal to $b$.
Let $\mathcal{I} \subset \Theta$ represent the set of internal branches (i.e.\ branches that connect two internal nodes) and $\mathcal{E} = \Theta \setminus \mathcal{I}$ represent the set of terminal branches (i.e.\ branches that connect an internal node to a tip node).
Let $\mathbf{D} = (D_1, ..., D_n)$ denote the trait values observed at the $n$ tips of $\tau$, $\mathbf{D}_{1:L} = \{ \mathbf{D}_1, ..., \mathbf{D}_L \}$ signify an alignment of length $L$, and $\mathbf{i} = (i_1, ..., i_{n-1})$ represent the unobserved internal node states of $\tau$.
In addition, we let $\mathbf{i}_b$ be the vector of internal node states for the subtree strictly beneath branch $b$.
Note that the internal nodes of $\tau$ are labeled with integers $\{ 1, ..., n-1 \}$ starting from the root of the tree; the corresponding labels of the branches and tips of $\tau$ are assigned arbitrarily.
We also introduce $\mathbf{i}^* = (i^*_1, ..., i^*_{n-1}, i^*_n, ..., i^*_{2n-1}) = (i_1, ..., i_{n-1}, D_1, ..., D_n)$, which is the concatenation of $\mathbf{i}$ and $\mathbf{D}$.
For each branch $b \in \Theta$, $p(b)$ and $c(b)$ represent the node labels (in $\mathbf{i}^*$) of the parent and child of branch $b$, respectively.
\par
Most stochastic mapping applications infer properties about the evolutionary process on the phylogeny through the use of a summary measure $H$.
We restrict attention to additive mapping summaries of the form:
\begin{equation}
H_\Omega \equiv H_\Omega(\mathbf{M}) = \sum_{b \in \Omega} h(\{ X_{bt} \}),
\label{eq1}
\end{equation}
where $\mathbf{M} = (\{ X_{1t} \}, ..., \{ X_{B_n t} \})$ denotes the collection of CTMC trajectories along the branches of $\tau$, $\Omega \subseteq \Theta$ represents a predefined set of branch indices, and $h$ signifies a summary measure applied to a single CTMC trajectory.
Let $\mathcal{L} \subset \{ 1, ..., m \}^2$ be a set of state pairs that labels substitutions of trait $X$ and $\mathbf{w} \subset \{ 0,1 \}^m$ be a set that labels states of trait $X$.
For any given CTMC path $\{ X_{t} \}$ in $[0,t)$, the two most popular choices of $h$ are $h_1(\{ X_{t} \})$, which counts the number of substitutions labeled by set $\mathcal{L}$, and $h_2(\{ X_{t} \})$, which measures the dwelling time in states labeled by set $\mathbf{w}$ \citep{minin2008fast}.
In this paper, we work exclusively with the summary function $h_1$ as this summary measure is used in both of our scientific applications.
However, we do note that our algorithmic results hold true regardless of the specific summary measure used.
\par
\citet{minin2008fast} were able to calculate the posterior mapping expectation $\text{E}(H_\Omega | \mathbf{D})$ for both $h_1$ and $h_2$ in $O(n)$ time and with $O(n)$ storage but were unable to achieve the same space-time complexity when calculating the posterior mapping variance $\text{Var}(H_\Omega | \mathbf{D})$.
Before we delve into the difficulties associated with computing $\text{Var}(H_\Omega | \mathbf{D})$, we refresh our readers on two important quantities:
\begin{gather}
\text{E}(H_\Omega^2 \mathbbm{1}_{\mathbf{D}}) = \text{E}(H_\Omega^2 | \mathbf{D}) \times \text{P}(\mathbf{D}),
\label{eq2} \\[5pt]
e^{[k]}_{ij}(h, t) = \text{E}\Bigl\{h(\{ X_{t} \})\bigl[h(\{ X_{t} \})-1\bigr]...\bigl[h(\{ X_{t} \})-k+1\bigr] \mathbbm{1}_{\{ X_{t} = j \}} {\Big |} X_0 = i\Bigr\},
\label{eq3}
\end{gather}
where $\mathbbm{1}_{\{ \cdot \}}$ represents the indicator function; $k = 1, 2, ...$; $i,j = 1, ..., m$; and $\text{P}(\mathbf{D})$ denotes the phylogenetic likelihood defined as the probability of observing the tip sequence $\mathbf{D}$.
Equation \eqref{eq2} connects the restricted mapping second moment $\text{E}(H_\Omega^2 \mathbbm{1}_{\mathbf{D}})$ to the posterior mapping second moment $\text{E}(H_\Omega^2 | \mathbf{D})$.
As \citet{minin2008fast} state, the restricted expectation in equation \eqref{eq2} integrates over all evolutionary mappings consistent with $\mathbf{D}$ on the tips of $\tau$.
Since $\text{P}(\mathbf{D})$ can be easily computed using the pruning algorithm \citep{felsenstein1981evolutionary}, we focus our attention on calculating $\text{E}(H_\Omega^2 \mathbbm{1}_{\mathbf{D}})$.
Quantity $e^{[k]}_{ij}(h, t)$ denotes the $k$th restricted factorial moment of $h(\{ X_{t} \})$ for a CTMC path $\{ X_{t} \}$ in $[0,t)$ that starts in state $i$ and ends in state $j$.
We let $\mathbf{e}^{[k]}(h, t) = \{ e^{[k]}_{ij}(h, t) \}$ represent the corresponding restricted factorial moment matrix.
\citet{minin2008counting} derive a simple recurrence relation to calculate $\mathbf{e}^{[k]}(h_1, t)$ for $k = 1, 2, ...$; a similar relation exists for $h_2$ as well \citep{minin2008fast}.
\par
To help us illustrate the computational challenges inherent in calculating $\text{Var}(H_\Omega | \mathbf{D})$, we express $\text{E}(H_\Omega^2 \mathbbm{1}_{\mathbf{D}})$ in the following manner (suppressing the fact that $b,b' \in \Omega$ for brevity):
\begin{align}
&\text{E}(H_\Omega^2 \mathbbm{1}_{\mathbf{D}}) = \text{E}\Biggl[\biggl(\sum_b h(\{ X_{bt} \})\biggr)^2 \mathbbm{1}_{\mathbf{D}}\Biggr]
\label{eq4} \\[15pt]
&= \sum_b \text{E}\bigl[h(\{ X_{bt} \})^2 \mathbbm{1}_{\mathbf{D}}\bigr] + \sum_{b \neq b'} \text{E}\bigl[h(\{ X_{bt} \}) h(\{ X_{b't} \}) \mathbbm{1}_{\mathbf{D}}\bigr]
\label{eq5} \\[15pt]
&= \sum_b \sum_{\mathbf{i}} \text{E}\bigl[h(\{ X_{bt} \})^2 | \ \mathbf{i},\mathbf{D}\bigr] \text{P}(\mathbf{i},\mathbf{D}) + \sum_{b \neq b'} \sum_{\mathbf{i}} \text{E}\bigl[h(\{ X_{bt} \}) h(\{ X_{b't} \}) | \ \mathbf{i},\mathbf{D}\bigr] \text{P}(\mathbf{i},\mathbf{D})
\label{eq6} \\[15pt]
\begin{split}
&= \sum_b \sum_{\mathbf{i}} \text{E}\bigl[h(\{ X_{bt} \})^2 | \ i^*_{p(b)}, i^*_{c(b)}\bigr] \pi_{i^*_1} \prod_{b^* \in \Theta} p_{i^*_{p(b^*)} i^*_{c(b^*)}}(t_{b^*}) \\
&+ \sum_{b \neq b'} \sum_{\mathbf{i}} \text{E}\bigl[h(\{ X_{bt} \}) h(\{ X_{b't} \}) | \ i^*_{p(b)}, i^*_{c(b)}, i^*_{p(b')}, i^*_{c(b')}\bigr] \pi_{i^*_1} \prod_{b^* \in \Theta} p_{i^*_{p(b^*)} i^*_{c(b^*)}}(t_{b^*})
\end{split}
\label{eq7} \\[15pt]
&= \sum_b \sum_{\mathbf{i}} \Bigl[e^{[2]}_{i^*_{p(b)} i^*_{c(b)}}(h, t_b) + e^{[1]}_{i^*_{p(b)} i^*_{c(b)}}(h, t_b)\Bigr] \pi_{i^*_1} \prod_{b^* \in \Theta \setminus \{ b \}} p_{i^*_{p(b^*)} i^*_{c(b^*)}}(t_{b^*})
\label{eq8} \\
&+ \sum_{b \neq b'} \sum_{\mathbf{i}} e^{[1]}_{i^*_{p(b)} i^*_{c(b)}}(h, t_b) e^{[1]}_{i^*_{p(b')} i^*_{c(b')}}(h, t_{b'}) \pi_{i^*_1} \prod_{b^* \in \Theta \setminus \{ b,b' \}} p_{i^*_{p(b^*)} i^*_{c(b^*)}}(t_{b^*}).
\label{eq9}
\end{align}
The single summation over $b$ in \eqref{eq8} can be efficiently computed by utilizing a modified version of the generalized pruning algorithm presented in \citep{minin2008fast}.
However, the straightforward double summation over $b \neq b'$ in \eqref{eq9} requires at most $O(B_n^2) = O((2n-2)^2) = O(n^2)$ computations.
In the next section, we describe how to overcome this computational roadblock via a post-order tree traversal algorithm that computes $\text{E}(H_\Omega^2 \mathbbm{1}_{\mathbf{D}})$ (and therefore $\text{Var}(H_\Omega | \mathbf{D})$) in $O(n)$ time and with $O(n)$ storage.

\section{Methods}

\subsection{Algorithm Setup}

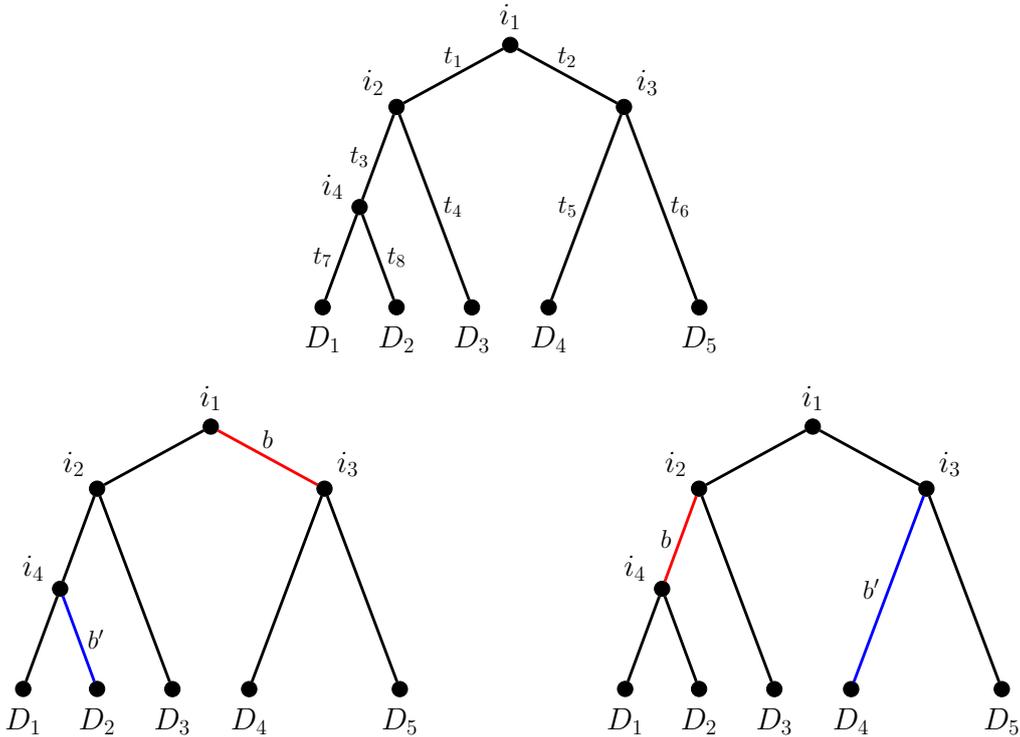
\begin{figure}[!ht]
\minipage{\textwidth}
\centering
\resizebox{!}{0.3\textwidth}{
\begin{tikzpicture}
  \node[circnode, label={[label distance=0.0325cm]90:{\LARGE $i_1$}}] (i1) {};
  \node[circnode, label={[label distance=0.01cm]135:{\LARGE $i_2$}}] (i2) [below left=1cm and 2cm of i1] {};
  \node[circnode, label={[label distance=0.01cm]45:{\LARGE $i_3$}}] (i3) [below right=1cm and 2cm of i1] {};
  \node[circnode, label={[label distance=1pt]170:{\LARGE $i_4$}}] (i4) [below left=1.75cm and 0.5cm of i2] {};
  \node[circnode, label={[label distance=0.1cm]270:{\LARGE $D_3$}}] (d3) [below right=3.725cm and 1.25cm of i2] {};
  \node[circnode, label={[label distance=0.1cm]270:{\LARGE $D_4$}}] (d4) [below left=3.725cm and 1.25cm of i3] {};
  \node[circnode, label={[label distance=0.1cm]270:{\LARGE $D_5$}}] (d5) [below right=3.725cm and 1.25cm of i3] {};
  \node[circnode, label={[label distance=0.1cm]270:{\LARGE $D_1$}}] (d1) [below left=1.75cm and 0.5cm of i4] {};
  \node[circnode, label={[label distance=0.1cm]270:{\LARGE $D_2$}}] (d2) [below right=1.75cm and 0.5cm of i4] {};
  \path[ultra thick] (i1) edge node[above=0.0325cm] {\Large $t_1$} (i2);
  \path[ultra thick] (i1) edge node[above=0.0325cm] {\Large $t_2$} (i3);
  \path[ultra thick] (i2) edge node[left=0.0325cm] {\Large $t_3$} (i4);
  \path[ultra thick] (i2) edge node[right=0.0325cm] {\Large $t_4$} (d3);
  \path[ultra thick] (i3) edge node[left=0.0325cm] {\Large $t_5$} (d4);
  \path[ultra thick] (i3) edge node[right=0.0325cm] {\Large $t_6$} (d5);
  \path[ultra thick] (i4) edge node[left=0.0325cm] {\Large $t_7$} (d1);
  \path[ultra thick] (i4) edge node[right=0.0325cm] {\Large $t_8$} (d2);
\end{tikzpicture}
}
\endminipage
\vspace{2.5mm}
\minipage{0.5\textwidth}
\centering
\resizebox{!}{0.6\textwidth}{
\begin{tikzpicture}
  \node[circnode, label={[label distance=0.0325cm]90:{\LARGE $i_1$}}] (i1) {};
  \node[circnode, label={[label distance=0.01cm]135:{\LARGE $i_2$}}] (i2) [below left=1cm and 2cm of i1] {};
  \node[circnode, label={[label distance=0.01cm]45:{\LARGE $i_3$}}] (i3) [below right=1cm and 2cm of i1] {};
  \node[circnode, label={[label distance=1pt]170:{\LARGE $i_4$}}] (i4) [below left=1.75cm and 0.5cm of i2] {};
  \node[circnode, label={[label distance=0.1cm]270:{\LARGE $D_3$}}] (d3) [below right=3.725cm and 1.25cm of i2] {};
  \node[circnode, label={[label distance=0.1cm]270:{\LARGE $D_4$}}] (d4) [below left=3.725cm and 1.25cm of i3] {};
  \node[circnode, label={[label distance=0.1cm]270:{\LARGE $D_5$}}] (d5) [below right=3.725cm and 1.25cm of i3] {};
  \node[circnode, label={[label distance=0.1cm]270:{\LARGE $D_1$}}] (d1) [below left=1.75cm and 0.5cm of i4] {};
  \node[circnode, label={[label distance=0.1cm]270:{\LARGE $D_2$}}] (d2) [below right=1.75cm and 0.5cm of i4] {};
  \path[ultra thick] (i1) edge node[above=0.0325cm] {} (i2);
  \path[ultra thick] (i1) edge[red] node[above=0.0325cm, black] {\Large $b$} (i3);
  \path[ultra thick] (i2) edge node[left=0.0325cm] {} (i4);
  \path[ultra thick] (i2) edge node[right=0.0325cm] {} (d3);
  \path[ultra thick] (i3) edge node[left=0.0325cm] {} (d4);
  \path[ultra thick] (i3) edge node[right=0.0325cm] {} (d5);
  \path[ultra thick] (i4) edge node[left=0.0325cm] {} (d1);
  \path[ultra thick] (i4) edge[blue] node[right=0.0325cm, black] {\Large $b'$} (d2);
\end{tikzpicture}}
\endminipage
\hfill
\minipage{0.5\textwidth}
\centering
\resizebox{!}{0.6\textwidth}{
\begin{tikzpicture}
  \node[circnode, label={[label distance=0.0325cm]90:{\LARGE $i_1$}}] (i1) {};
  \node[circnode, label={[label distance=0.01cm]135:{\LARGE $i_2$}}] (i2) [below left=1cm and 2cm of i1] {};
  \node[circnode, label={[label distance=0.01cm]45:{\LARGE $i_3$}}] (i3) [below right=1cm and 2cm of i1] {};
  \node[circnode, label={[label distance=1pt]170:{\LARGE $i_4$}}] (i4) [below left=1.75cm and 0.5cm of i2] {};
  \node[circnode, label={[label distance=0.1cm]270:{\LARGE $D_3$}}] (d3) [below right=3.725cm and 1.25cm of i2] {};
  \node[circnode, label={[label distance=0.1cm]270:{\LARGE $D_4$}}] (d4) [below left=3.725cm and 1.25cm of i3] {};
  \node[circnode, label={[label distance=0.1cm]270:{\LARGE $D_5$}}] (d5) [below right=3.725cm and 1.25cm of i3] {};
  \node[circnode, label={[label distance=0.1cm]270:{\LARGE $D_1$}}] (d1) [below left=1.75cm and 0.5cm of i4] {};
  \node[circnode, label={[label distance=0.1cm]270:{\LARGE $D_2$}}] (d2) [below right=1.75cm and 0.5cm of i4] {};
  \path[ultra thick] (i1) edge node[above=0.0325cm] {} (i2);
  \path[ultra thick] (i1) edge node[above=0.0325cm] {} (i3);
  \path[ultra thick] (i2) edge[red] node[left=0.0325cm, black] {\Large $b$} (i4);
  \path[ultra thick] (i2) edge node[right=0.0325cm] {} (d3);
  \path[ultra thick] (i3) edge[blue] node[left=0.0325cm, black] {\Large $b'$} (d4);
  \path[ultra thick] (i3) edge node[right=0.0325cm] {} (d5);
  \path[ultra thick] (i4) edge node[left=0.0325cm] {} (d1);
  \path[ultra thick] (i4) edge node[right=0.0325cm] {} (d2);
\end{tikzpicture}
}
\endminipage
\caption{The example phylogenies we use to help motivate our algorithmic procedure.  (Top) The ``uncolored'' tree used to create all the ``colored'' phylogenies.  This tree has $n=5$ tips, $B_n=8$ branches with branch lengths $\mathbf{t} = (t_1, ..., t_8)$, internal node states $\mathbf{i} = (i_1, ..., i_4)$, and tip states $\mathbf{D} = (D_1, ..., D_5)$.  In addition, $\Theta = \{ 1, 2, 3, 4, 5, 6, 7, 8 \}$, $\mathcal{I} = \{ 1, 2, 3 \}$, and $\mathcal{E} = \{ 4, 5, 6, 7, 8 \}$.  (Bottom) Two ``colored'' phylogenies that illustrate the calculation of $\text{E}\bigl[h(\{ X_{bt} \}) h(\{ X_{b't} \}) \mathbbm{1}_{\mathbf{D}}\bigr]$.  The colored branches specify the locations of the restricted first moments, while the uncolored branches determine the locations of the transition probabilities.  The first tree (left) and second tree (right) visualize the calculations of $\text{E}\bigl[h(\{ X_{2t} \}) h(\{ X_{8t} \}) \mathbbm{1}_{\mathbf{D}}\bigr]$ and $\text{E}\bigl[h(\{ X_{3t} \}) h(\{ X_{5t} \}) \mathbbm{1}_{\mathbf{D}}\bigr]$, respectively, for the ``uncolored'' tree shown above.}
\label{extree}
\end{figure}

To help motivate our procedure and make it easier to understand, we utilize illustrations of various ``colored'' phylogenies.
\autoref*{extree} displays the example ``uncolored'' tree we use to create all these illustrations.
Note that for a given $b \neq b'$ where $b,b' \in \Omega$, the corresponding summand in \eqref{eq9} is:
\begin{equation}
\sum_{\mathbf{i}} e^{[1]}_{i^*_{p(b)} i^*_{c(b)}}(h, t_b) e^{[1]}_{i^*_{p(b')} i^*_{c(b')}}(h, t_{b'}) \pi_{i^*_1} \prod_{b^* \in \Theta \setminus \{ b,b' \}} p_{i^*_{p(b^*)} i^*_{c(b^*)}}(t_{b^*}).
\label{eq10}
\end{equation}
If we replace the restricted first moments in \eqref{eq10} with the appropriate transition probabilities, then expression \eqref{eq10} would represent the phylogenetic likelihood.
From equation \eqref{eq5}, we know that expression \eqref{eq10} is equal to the restricted product moment $\text{E}\bigl[h(\{ X_{bt} \}) h(\{ X_{b't} \}) \mathbbm{1}_{\mathbf{D}}\bigr]$.
In \autoref*{extree}, we present two ``colored'' phylogenies to help visualize the calculation of $\text{E}\bigl[h(\{ X_{bt} \}) h(\{ X_{b't} \}) \mathbbm{1}_{\mathbf{D}}\bigr]$.
Thus, we can visualize the double sum over $b \neq b'$ in \eqref{eq9} by imagining the red and blue colorings being permuted across all branches in $\Omega$.
The cached vectors used in our procedure are described in a similar fashion.
\par
Let $\mathbf{F}_u = (F_{u1}, ..., F_{um})^T$ be the vector of partial likelihoods at node $u$, where $F_{ui}$ denotes the probability of the observed data at only the tips that descend from node $u$, given that the state of node $u$ is $i$.
We let $\mathbf{S}_b = (S_{b1}, ..., S_{bm})^T$ denote the vector of directional likelihoods at branch $b$, where $S_{bi}$ represents the likelihood of the observed data at only the tips that descend from branch $b$, given that the state of parent node $p(b)$ is $i$.
\citet{minin2008fast} utilize these $\mathbf{F}_u$ and $\mathbf{S}_b$ vectors within their algorithm for computing the posterior mapping expectation $\text{E}(H_\Omega | \mathbf{D})$.
\par
We also define vectors $\mathbf{V}^{[1]}_b = \bigl(V^{[1]}_{b1}, ..., V^{[1]}_{bm}\bigr)^T$, $\mathbf{V}^{[2]}_b = \bigl(V^{[2]}_{b1}, ..., V^{[2]}_{bm}\bigr)^T$, and $\mathbf{W}_b = (W_{b1}, ..., W_{bm})^T$.
Elements of vector $\mathbf{V}^{[1]}_b$  are defined as follows:
\begin{equation}
V^{[1]}_{bi} = \sum_{b^\dagger} \sum_{\mathbf{i}_b} e^{[1]}_{i^*_{p(b^\dagger)} i^*_{c(b^\dagger)}}(h, t_{b^\dagger}) \prod_{b^* \in \Theta_b \setminus \{ b^\dagger \}} p_{i^*_{p(b^*)} i^*_{c(b^*)}}(t_{b^*}),
\label{eq11}
\end{equation}
where the state of parent node $p(b)$ is $i$ and $b^\dagger \in \Omega_b$ for the set of branches of interest in the subtree that is ``below'' branch $b$ (including $b$), $\Omega_b = \Omega \cap \Theta_b$.
An illustration of $\mathbf{V}^{[1]}_b$ is given in \autoref*{vwlist}.
We can interpret $\boldsymbol{\pi}^T \mathbf{V}^{[1]}_b$ as the restricted mapping first moment $\text{E}(H_{\Omega_b} \mathbbm{1}_{\mathbf{D}})$ for the subtree that is ``below'' branch $b$ (including $b$).
\citet{kenney2012hessian} cache a vector similar to $\mathbf{V}^{[1]}_b$ within their algorithm for computing second derivatives of phylogenetic likelihood functions.
If we replace the restricted first moment in expression \eqref{eq11} with the appropriate transition probability derivative, then we would recover this cached vector used in \citep{kenney2012hessian}.
Similarly, $V^{[2]}_{bi}$ is defined the same as $V^{[1]}_{bi}$, except in the case of $V^{[2]}_{bi}$, we replace the restricted first moment in expression \eqref{eq11} with the corresponding second restricted factorial moment $e^{[2]}_{i^*_{p(b^\dagger)} i^*_{c(b^\dagger)}}(h, t_{b^\dagger})$.
The visual depiction of $\mathbf{V}^{[2]}_b$ is analogous to that of $\mathbf{V}^{[1]}_b$ in \autoref*{vwlist}.
Elements of vector $\mathbf{W}_b$ are defined as follows:
\begin{equation}
W_{bi} = \sum_{b^\dagger \neq b^{\dagger \dagger}} \sum_{\mathbf{i}_b} e^{[1]}_{i^*_{p(b^\dagger)} i^*_{c(b^\dagger)}}(h, t_{b^\dagger}) e^{[1]}_{i^*_{p(b^{\dagger \dagger})} i^*_{c(b^{\dagger \dagger})}}(h, t_{b^{\dagger \dagger}}) \prod_{b^* \in \Theta_b \setminus \{ b^\dagger,b^{\dagger \dagger} \}} p_{i^*_{p(b^*)} i^*_{c(b^*)}}(t_{b^*}),
\label{eq12}
\end{equation}
where the state of parent node $p(b)$ is $i$ and $b^\dagger,b^{\dagger \dagger} \in \Omega_b$ for $\Omega_b$ defined as above.
A pictorial representation of $\mathbf{W}_b$ is provided in \autoref*{vwlist}.
$\boldsymbol{\pi}^T \mathbf{W}_b$ can be viewed as the sum of restricted product moments $\text{E}\bigl[h(\{ X_{b^\dagger t} \}) h(\{ X_{b^{\dagger \dagger} t} \}) \mathbbm{1}_{\mathbf{D}}\bigr]$ over $b^\dagger \neq b^{\dagger \dagger} \ (b^\dagger,b^{\dagger \dagger} \in \Omega_b)$ for the subtree that is ``below'' branch $b$ (including $b$).
In the next subsection, we describe our algorithm in full detail and provide some intuition behind the recursive formulas used in our procedure.
\begin{figure}
{\Large (A)}
\begin{mdframed}
\minipage{0.3\textwidth}
\centering
\resizebox{!}{\textwidth}{
\begin{tikzpicture}
  \node[circnode, label={[label distance=0.0325cm, xshift=0.6cm]90:{\LARGE $i_{p(b)} = i_1$}}] (i1) {};
  \node[circnode, label={[label distance=0.01cm]135:{\LARGE $i_2$}}] (i2) [below left=1cm and 2cm of i1] {};
  \node[circnode, label={[label distance=1pt]170:{\LARGE $i_4$}}] (i4) [below left=1.75cm and 0.5cm of i2] {};
  \node[circnode, label={[label distance=0.1cm]270:{\LARGE $D_3$}}] (d3) [below right=3.725cm and 1.25cm of i2] {};
  \node[circnode, label={[label distance=0.1cm]270:{\LARGE $D_1$}}] (d1) [below left=1.75cm and 0.5cm of i4] {};
  \node[circnode, label={[label distance=0.1cm]270:{\LARGE $D_2$}}] (d2) [below right=1.75cm and 0.5cm of i4] {};
  \path[ultra thick] (i1) edge[red] node[above=0.0325cm, black] {\Large $b^\dagger$} node[below right=0.005cm and 0.005cm, black] {\Large $b = 1$} (i2);
  \path[ultra thick] (i2) edge node[left=0.0325cm] {} (i4);
  \path[ultra thick] (i2) edge node[right=0.0325cm] {} (d3);
  \path[ultra thick] (i4) edge node[left=0.0325cm] {} (d1);
  \path[ultra thick] (i4) edge node[right=0.0325cm] {} (d2);
\end{tikzpicture}
}
\caption*{
\minipage{\textwidth}
\centering
\vspace{-5mm}
\footnotesize
\begin{align*}
\sum_{i_2,i_4} \Bigl[&e^{[1]}_{i_1 i_2}(h, t_1) p_{i_2 i_4}(t_3) p_{i_2 D_3}(t_4)\Bigr. \\[-10pt]
&\times \Bigl.p_{i_4 D_1}(t_7) p_{i_4 D_2}(t_8)\Bigr]
\end{align*}
\endminipage
}
\endminipage
{\Huge $+$}
\minipage{0.3\textwidth}
\centering
\resizebox{!}{\textwidth}{
\begin{tikzpicture}
  \node[circnode, label={[label distance=0.0325cm, xshift=0.6cm]90:{\LARGE $i_{p(b)} = i_1$}}] (i1) {};
  \node[circnode, label={[label distance=0.01cm]135:{\LARGE $i_2$}}] (i2) [below left=1cm and 2cm of i1] {};
  \node[circnode, label={[label distance=1pt]170:{\LARGE $i_4$}}] (i4) [below left=1.75cm and 0.5cm of i2] {};
  \node[circnode, label={[label distance=0.1cm]270:{\LARGE $D_3$}}] (d3) [below right=3.725cm and 1.25cm of i2] {};
  \node[circnode, label={[label distance=0.1cm]270:{\LARGE $D_1$}}] (d1) [below left=1.75cm and 0.5cm of i4] {};
  \node[circnode, label={[label distance=0.1cm]270:{\LARGE $D_2$}}] (d2) [below right=1.75cm and 0.5cm of i4] {};
  \path[ultra thick] (i1) edge node[below right=0.005cm and 0.005cm, black] {\Large $b = 1$} (i2);
  \path[ultra thick] (i2) edge[red] node[left=0.0325cm, black] {\Large $b^\dagger$} (i4);
  \path[ultra thick] (i2) edge node[right=0.0325cm] {} (d3);
  \path[ultra thick] (i4) edge node[left=0.0325cm] {} (d1);
  \path[ultra thick] (i4) edge node[right=0.0325cm] {} (d2);
\end{tikzpicture}
}
\caption*{
\minipage{\textwidth}
\centering
\vspace{-5mm}
\footnotesize
\begin{align*}
\sum_{i_2,i_4} \Bigl[&p_{i_1 i_2}(t_1) e^{[1]}_{i_2 i_4}(h, t_3) p_{i_2 D_3}(t_4)\Bigr. \\[-10pt]
&\times \Bigl.p_{i_4 D_1}(t_7) p_{i_4 D_2}(t_8)\Bigr]
\end{align*}
\endminipage
}
\endminipage
{\Huge $+$}
\minipage{0.3\textwidth}
\centering
\resizebox{!}{\textwidth}{
\begin{tikzpicture}
  \node[circnode, label={[label distance=0.0325cm, xshift=0.6cm]90:{\LARGE $i_{p(b)} = i_1$}}] (i1) {};
  \node[circnode, label={[label distance=0.01cm]135:{\LARGE $i_2$}}] (i2) [below left=1cm and 2cm of i1] {};
  \node[circnode, label={[label distance=1pt]170:{\LARGE $i_4$}}] (i4) [below left=1.75cm and 0.5cm of i2] {};
  \node[circnode, label={[label distance=0.1cm]270:{\LARGE $D_3$}}] (d3) [below right=3.725cm and 1.25cm of i2] {};
  \node[circnode, label={[label distance=0.1cm]270:{\LARGE $D_1$}}] (d1) [below left=1.75cm and 0.5cm of i4] {};
  \node[circnode, label={[label distance=0.1cm]270:{\LARGE $D_2$}}] (d2) [below right=1.75cm and 0.5cm of i4] {};
  \path[ultra thick] (i1) edge node[below right=0.005cm and 0.005cm, black] {\Large $b = 1$} (i2);
  \path[ultra thick] (i2) edge node[left=0.0325cm] {} (i4);
  \path[ultra thick] (i2) edge[red] node[right=0.0325cm, black] {\Large $b^\dagger$} (d3);
  \path[ultra thick] (i4) edge node[left=0.0325cm] {} (d1);
  \path[ultra thick] (i4) edge node[right=0.0325cm] {} (d2);
\end{tikzpicture}
}
\caption*{
\minipage{\textwidth}
\centering
\vspace{-5mm}
\footnotesize
\begin{align*}
\sum_{i_2,i_4} \Bigl[&p_{i_1 i_2}(t_1) p_{i_2 i_4}(t_3) e^{[1]}_{i_2 D_3}(h, t_4)\Bigr. \\[-10pt]
&\times \Bigl.p_{i_4 D_1}(t_7) p_{i_4 D_2}(t_8)\Bigr]
\end{align*}
\endminipage
}
\endminipage
\end{mdframed}
\vfill
{\Large (B)}
\begin{mdframed}
\minipage{0.3\textwidth}
\centering
\resizebox{!}{\textwidth}{
\begin{tikzpicture}
  \node[circnode, label={[label distance=0.0325cm, xshift=0.6cm]90:{\LARGE $i_{p(b)} = i_1$}}] (i1) {};
  \node[circnode, label={[label distance=0.01cm]135:{\LARGE $i_2$}}] (i2) [below left=1cm and 2cm of i1] {};
  \node[circnode, label={[label distance=1pt]170:{\LARGE $i_4$}}] (i4) [below left=1.75cm and 0.5cm of i2] {};
  \node[circnode, label={[label distance=0.1cm]270:{\LARGE $D_3$}}] (d3) [below right=3.725cm and 1.25cm of i2] {};
  \node[circnode, label={[label distance=0.1cm]270:{\LARGE $D_1$}}] (d1) [below left=1.75cm and 0.5cm of i4] {};
  \node[circnode, label={[label distance=0.1cm]270:{\LARGE $D_2$}}] (d2) [below right=1.75cm and 0.5cm of i4] {};
  \path[ultra thick] (i1) edge[red] node[above=0.0325cm, black] {\Large $b^\dagger$} node[below right=0.005cm and 0.005cm, black] {\Large $b = 1$} (i2);
  \path[ultra thick] (i2) edge[blue] node[left=0.0325cm, black] {\Large $b^{\dagger \dagger}$} (i4);
  \path[ultra thick] (i2) edge node[right=0.0325cm] {} (d3);
  \path[ultra thick] (i4) edge node[left=0.0325cm] {} (d1);
  \path[ultra thick] (i4) edge node[right=0.0325cm] {} (d2);
\end{tikzpicture}
}
\caption*{
\minipage{\textwidth}
\centering
\vspace{-5mm}
\footnotesize
\begin{align*}
\sum_{i_2,i_4} \Bigl[&e^{[1]}_{i_1 i_2}(h, t_1) e^{[1]}_{i_2 i_4}(h, t_3) p_{i_2 D_3}(t_4)\Bigr. \\[-10pt]
&\times \Bigl.p_{i_4 D_1}(t_7) p_{i_4 D_2}(t_8)\Bigr]
\end{align*}
\endminipage
}
\endminipage
{\Huge $+$}
\minipage{0.3\textwidth}
\centering
\resizebox{!}{\textwidth}{
\begin{tikzpicture}
  \node[circnode, label={[label distance=0.0325cm, xshift=0.6cm]90:{\LARGE $i_{p(b)} = i_1$}}] (i1) {};
  \node[circnode, label={[label distance=0.01cm]135:{\LARGE $i_2$}}] (i2) [below left=1cm and 2cm of i1] {};
  \node[circnode, label={[label distance=1pt]170:{\LARGE $i_4$}}] (i4) [below left=1.75cm and 0.5cm of i2] {};
  \node[circnode, label={[label distance=0.1cm]270:{\LARGE $D_3$}}] (d3) [below right=3.725cm and 1.25cm of i2] {};
  \node[circnode, label={[label distance=0.1cm]270:{\LARGE $D_1$}}] (d1) [below left=1.75cm and 0.5cm of i4] {};
  \node[circnode, label={[label distance=0.1cm]270:{\LARGE $D_2$}}] (d2) [below right=1.75cm and 0.5cm of i4] {};
  \path[ultra thick] (i1) edge node[below right=0.005cm and 0.005cm, black] {\Large $b = 1$} (i2);
  \path[ultra thick] (i2) edge[red] node[left=0.0325cm, black] {\Large $b^\dagger$} (i4);
  \path[ultra thick] (i2) edge[blue] node[right=0.0325cm, black] {\Large $b^{\dagger \dagger}$} (d3);
  \path[ultra thick] (i4) edge node[left=0.0325cm] {} (d1);
  \path[ultra thick] (i4) edge node[right=0.0325cm] {} (d2);
\end{tikzpicture}
}
\caption*{
\minipage{\textwidth}
\centering
\vspace{-5mm}
\footnotesize
\begin{align*}
\sum_{i_2,i_4} \Bigl[&p_{i_1 i_2}(t_1) e^{[1]}_{i_2 i_4}(h, t_3) e^{[1]}_{i_2 D_3}(h, t_4)\Bigr. \\[-10pt]
&\times \Bigl.p_{i_4 D_1}(t_7) p_{i_4 D_2}(t_8)\Bigr]
\end{align*}
\endminipage
}
\endminipage
{\Huge $+$}
\minipage{0.3\textwidth}
\centering
\resizebox{!}{\textwidth}{
\begin{tikzpicture}
  \node[circnode, label={[label distance=0.0325cm, xshift=0.6cm]90:{\LARGE $i_{p(b)} = i_1$}}] (i1) {};
  \node[circnode, label={[label distance=0.01cm]135:{\LARGE $i_2$}}] (i2) [below left=1cm and 2cm of i1] {};
  \node[circnode, label={[label distance=1pt]170:{\LARGE $i_4$}}] (i4) [below left=1.75cm and 0.5cm of i2] {};
  \node[circnode, label={[label distance=0.1cm]270:{\LARGE $D_3$}}] (d3) [below right=3.725cm and 1.25cm of i2] {};
  \node[circnode, label={[label distance=0.1cm]270:{\LARGE $D_1$}}] (d1) [below left=1.75cm and 0.5cm of i4] {};
  \node[circnode, label={[label distance=0.1cm]270:{\LARGE $D_2$}}] (d2) [below right=1.75cm and 0.5cm of i4] {};
  \path[ultra thick] (i1) edge[blue] node[above=0.0325cm, black] {\Large $b^{\dagger \dagger}$} node[below right=0.005cm and 0.005cm, black] {\Large $b = 1$} (i2);
  \path[ultra thick] (i2) edge node[left=0.0325cm] {} (i4);
  \path[ultra thick] (i2) edge[red] node[right=0.0325cm, black] {\Large $b^\dagger$} (d3);
  \path[ultra thick] (i4) edge node[left=0.0325cm] {} (d1);
  \path[ultra thick] (i4) edge node[right=0.0325cm] {} (d2);
\end{tikzpicture}
}
\caption*{
\minipage{\textwidth}
\centering
\vspace{-5mm}
\footnotesize
\begin{align*}
\sum_{i_2,i_4} \Bigl[&e^{[1]}_{i_1 i_2}(h, t_1) p_{i_2 i_4}(t_3) e^{[1]}_{i_2 D_3}(h, t_4)\Bigr. \\[-10pt]
&\times \Bigl.p_{i_4 D_1}(t_7) p_{i_4 D_2}(t_8)\Bigr]
\end{align*}
\endminipage
}
\endminipage
\vspace{2.5mm}
\minipage{0.3\textwidth}
\centering
{\Huge $+$}
\endminipage
\phantom{\Huge $+$}
\minipage{0.3\textwidth}
\centering
{\Huge $+$}
\endminipage
\phantom{\Huge $+$}
\minipage{0.3\textwidth}
\centering
{\Huge $+$}
\endminipage
\vspace{2.5mm}
\minipage{0.3\textwidth}
\centering
\resizebox{!}{\textwidth}{
\begin{tikzpicture}
  \node[circnode, label={[label distance=0.0325cm, xshift=0.6cm]90:{\LARGE $i_{p(b)} = i_1$}}] (i1) {};
  \node[circnode, label={[label distance=0.01cm]135:{\LARGE $i_2$}}] (i2) [below left=1cm and 2cm of i1] {};
  \node[circnode, label={[label distance=1pt]170:{\LARGE $i_4$}}] (i4) [below left=1.75cm and 0.5cm of i2] {};
  \node[circnode, label={[label distance=0.1cm]270:{\LARGE $D_3$}}] (d3) [below right=3.725cm and 1.25cm of i2] {};
  \node[circnode, label={[label distance=0.1cm]270:{\LARGE $D_1$}}] (d1) [below left=1.75cm and 0.5cm of i4] {};
  \node[circnode, label={[label distance=0.1cm]270:{\LARGE $D_2$}}] (d2) [below right=1.75cm and 0.5cm of i4] {};
  \path[ultra thick] (i1) edge[blue] node[above=0.0325cm, black] {\Large $b^{\dagger \dagger}$} node[below right=0.005cm and 0.005cm, black] {\Large $b = 1$} (i2);
  \path[ultra thick] (i2) edge[red] node[left=0.0325cm, black] {\Large $b^\dagger$} (i4);
  \path[ultra thick] (i2) edge node[right=0.0325cm] {} (d3);
  \path[ultra thick] (i4) edge node[left=0.0325cm] {} (d1);
  \path[ultra thick] (i4) edge node[right=0.0325cm] {} (d2);
\end{tikzpicture}
}
\caption*{
\minipage{\textwidth}
\centering
\vspace{-5mm}
\footnotesize
\begin{align*}
\sum_{i_2,i_4} \Bigl[&e^{[1]}_{i_1 i_2}(h, t_1) e^{[1]}_{i_2 i_4}(h, t_3) p_{i_2 D_3}(t_4)\Bigr. \\[-10pt]
&\times \Bigl.p_{i_4 D_1}(t_7) p_{i_4 D_2}(t_8)\Bigr]
\end{align*}
\endminipage
}
\endminipage
{\Huge $+$}
\minipage{0.3\textwidth}
\centering
\resizebox{!}{\textwidth}{
\begin{tikzpicture}
  \node[circnode, label={[label distance=0.0325cm, xshift=0.6cm]90:{\LARGE $i_{p(b)} = i_1$}}] (i1) {};
  \node[circnode, label={[label distance=0.01cm]135:{\LARGE $i_2$}}] (i2) [below left=1cm and 2cm of i1] {};
  \node[circnode, label={[label distance=1pt]170:{\LARGE $i_4$}}] (i4) [below left=1.75cm and 0.5cm of i2] {};
  \node[circnode, label={[label distance=0.1cm]270:{\LARGE $D_3$}}] (d3) [below right=3.725cm and 1.25cm of i2] {};
  \node[circnode, label={[label distance=0.1cm]270:{\LARGE $D_1$}}] (d1) [below left=1.75cm and 0.5cm of i4] {};
  \node[circnode, label={[label distance=0.1cm]270:{\LARGE $D_2$}}] (d2) [below right=1.75cm and 0.5cm of i4] {};
  \path[ultra thick] (i1) edge node[below right=0.005cm and 0.005cm, black] {\Large $b = 1$} (i2);
  \path[ultra thick] (i2) edge[blue] node[left=0.0325cm, black] {\Large $b^{\dagger \dagger}$} (i4);
  \path[ultra thick] (i2) edge[red] node[right=0.0325cm, black] {\Large $b^\dagger$} (d3);
  \path[ultra thick] (i4) edge node[left=0.0325cm] {} (d1);
  \path[ultra thick] (i4) edge node[right=0.0325cm] {} (d2);
\end{tikzpicture}
}
\caption*{
\minipage{\textwidth}
\centering
\vspace{-5mm}
\footnotesize
\begin{align*}
\sum_{i_2,i_4} \Bigl[&p_{i_1 i_2}(t_1) e^{[1]}_{i_2 i_4}(h, t_3) e^{[1]}_{i_2 D_3}(h, t_4)\Bigr. \\[-10pt]
&\times \Bigl.p_{i_4 D_1}(t_7) p_{i_4 D_2}(t_8)\Bigr]
\end{align*}
\endminipage
}
\endminipage
{\Huge $+$}
\minipage{0.3\textwidth}
\centering
\resizebox{!}{\textwidth}{
\begin{tikzpicture}
  \node[circnode, label={[label distance=0.0325cm, xshift=0.6cm]90:{\LARGE $i_{p(b)} = i_1$}}] (i1) {};
  \node[circnode, label={[label distance=0.01cm]135:{\LARGE $i_2$}}] (i2) [below left=1cm and 2cm of i1] {};
  \node[circnode, label={[label distance=1pt]170:{\LARGE $i_4$}}] (i4) [below left=1.75cm and 0.5cm of i2] {};
  \node[circnode, label={[label distance=0.1cm]270:{\LARGE $D_3$}}] (d3) [below right=3.725cm and 1.25cm of i2] {};
  \node[circnode, label={[label distance=0.1cm]270:{\LARGE $D_1$}}] (d1) [below left=1.75cm and 0.5cm of i4] {};
  \node[circnode, label={[label distance=0.1cm]270:{\LARGE $D_2$}}] (d2) [below right=1.75cm and 0.5cm of i4] {};
  \path[ultra thick] (i1) edge[red] node[above=0.0325cm, black] {\Large $b^\dagger$} node[below right=0.005cm and 0.005cm, black] {\Large $b = 1$} (i2);
  \path[ultra thick] (i2) edge node[left=0.0325cm] {} (i4);
  \path[ultra thick] (i2) edge[blue] node[right=0.0325cm, black] {\Large $b^{\dagger \dagger}$} (d3);
  \path[ultra thick] (i4) edge node[left=0.0325cm] {} (d1);
  \path[ultra thick] (i4) edge node[right=0.0325cm] {} (d2);
\end{tikzpicture}
}
\caption*{
\minipage{\textwidth}
\centering
\vspace{-5mm}
\footnotesize
\begin{align*}
\sum_{i_2,i_4} \Bigl[&e^{[1]}_{i_1 i_2}(h, t_1) p_{i_2 i_4}(t_3) e^{[1]}_{i_2 D_3}(h, t_4)\Bigr. \\[-10pt]
&\times \Bigl.p_{i_4 D_1}(t_7) p_{i_4 D_2}(t_8)\Bigr]
\end{align*}
\endminipage
}
\endminipage
\end{mdframed}
\caption{Visual depictions of the $\mathbf{V}^{[1]}_b$ and $\mathbf{W}_b$ vectors.  (A) An illustration of the $\mathbf{V}^{[1]}_b$ vector.  $V^{[1]}_{bi}$ can be interpreted as the sum over all ``single-colored'' phylogenies for the subtree defined by $\Theta_b$ and the predefined set of ``colored'' branches $\Omega_b = \Omega \cap \Theta_b$, conditional on the state of parent node $p(b)$ being $i$.  We illustrate $\mathbf{V}^{[1]}_1$ for the ``uncolored'' phylogeny given in \autoref*{extree}, where $\Omega_1 = \{ 1, 3, 4 \}$.  (B) An illustration of the $\mathbf{W}_b$ vector.  $W_{bi}$ can be interpreted as the sum over all ``double-colored'' phylogenies for the same subtree and set of ``colored'' branches as described in (A), conditional on the state of parent node $p(b)$ being $i$.  We illustrate $\mathbf{W}_1$ for the ``uncolored'' tree displayed in \autoref*{extree}, where $\Omega_1 = \{ 1, 3, 4 \}$.}
\label{vwlist}
\end{figure}

\subsection{Algorithm Recursion}

Our post-order tree traversal algorithm recursively computes $\mathbf{F}_u$, $\mathbf{S}_b$, $\mathbf{V}^{[1]}_b$, $\mathbf{V}^{[2]}_b$, and $\mathbf{W}_b$ at all nodes $u \in \{ 1, ..., n-1, n, ..., 2n-1 \}$ and branches $b \in \Theta$.
Like the pruning algorithm, this procedure starts at the tips of the tree and continues through all ancestral nodes until it arrives at the root of the tree.
We start by describing how these vectors are initialized at the terminal nodes/branches of $\tau$ and then specify the recursive formulas used to calculate $\mathbf{F}_u$, $\mathbf{S}_b$, $\mathbf{V}^{[1]}_b$, $\mathbf{V}^{[2]}_b$, and $\mathbf{W}_b$ at the internal nodes/branches of $\tau$.
\par
First, we follow standard practice and set $F_{ui} = \mathbbm{1}_{\{ i = i^*_u \}}$ for all terminal nodes $u \in \{ n, ..., 2n-1 \}$ and $i = 1, ..., m$.
Oftentimes, we have partially observed and/or missing data at the tips of $\tau$ and our initialization of $F_{ui}$ can be adjusted to reflect this information \citep{felsenstein1981evolutionary}.
For terminal branches $b \in \mathcal{E}$, we set:
\begin{equation}
S_{bi} = \sum_{j=1}^m p_{ij}(t_b) F_{c(b) j},
\label{eq13}
\end{equation}
for $i = 1, ..., m$.
Using matrix notation, we express the equation in \eqref{eq13} as $\mathbf{S}_b = \mathbf{P}(t_b) \mathbf{F}_{c(b)}$.
The initializations of $\mathbf{V}^{[1]}_b$ and $\mathbf{V}^{[2]}_b$ depend on whether or not $b \in \Omega$.
For all terminal branches $b \in \mathcal{E}$, we define:
\begin{gather}
V^{[1]}_{bi} = \sum_{j=1}^m e^{[1]}_{ij}(h, t_b) F_{c(b) j} \mathbbm{1}_{\{ b \in \Omega \}},
\label{eq14} \\[10pt]
V^{[2]}_{bi} = \sum_{j=1}^m e^{[2]}_{ij}(h, t_b) F_{c(b) j} \mathbbm{1}_{\{ b \in \Omega \}},
\label{eq15}
\end{gather}
for $i = 1, ..., m$.
The vectorized representations of equations \eqref{eq14} and \eqref{eq15} are $\mathbf{V}^{[1]}_b = \mathbf{e}^{[1]}(h, t_b) \mathbf{F}_{c(b)} \mathbbm{1}_{\{ b \in \Omega \}}$ and $\mathbf{V}^{[2]}_b = \mathbf{e}^{[2]}(h, t_b) \mathbf{F}_{c(b)} \mathbbm{1}_{\{ b \in \Omega \}}$, respectively.
Finally, $W_{bi} = 0$ for all terminal branches $b \in \mathcal{E}$ and $i = 1, ..., m$.
Note that the definitions of $\mathbf{V}^{[1]}_b$, $\mathbf{V}^{[2]}_b$, and $\mathbf{W}_b$ for $b \in \mathcal{E}$ are consistent with the illustrations provided in \autoref*{vwlist}.
\par
Now, we present the recursive formulas that compute $\mathbf{F}_u$, $\mathbf{S}_b$, $\mathbf{V}^{[1]}_b$, $\mathbf{V}^{[2]}_b$, and $\mathbf{W}_b$ for internal nodes $u \in \{ 1, ..., n-1 \}$ and internal branches $b \in \mathcal{I}$.
The recursion for the partial likelihood $F_{ui}$ is the centerpiece of Felsenstein's pruning algorithm \citep{felsenstein1981evolutionary}:
\begin{equation}
F_{ui} = \underbrace{\left[\sum_{j=1}^m p_{ij}(t_{b_1}) F_{c(b_1) j}\right]}_{S_{b_1 i}} \times \underbrace{\left[\sum_{j=1}^m p_{ij}(t_{b_2}) F_{c(b_2) j}\right]}_{S_{b_2 i}},
\label{eq16}
\end{equation}
where $b_1$ and $b_2$ represent the two branches connecting node $u$ to its two child nodes and $i = 1, ..., m$.
In addition, as shown in the brackets above, equation \eqref{eq13} also denotes the recursion for $S_{bi}$ at internal branches $b \in \mathcal{I}$ \citep{felsenstein1981evolutionary}.
Thus, the recursive formula for $F_{ui}$ in \eqref{eq16} can be compactly expressed as $\mathbf{F}_u = \mathbf{S}_{b_1} \circ \mathbf{S}_{b_2}$, where $\circ$ symbolizes element-wise multiplication between two vectors.
The recursive equations used to calculate $V^{[1]}_{bi}$, $V^{[2]}_{bi}$, and $W_{bi}$ at internal branches $b \in \mathcal{I}$ are:
\begin{gather}
V^{[1]}_{bi} = \sum_{j=1}^m \biggl[e^{[1]}_{ij}(h, t_b) F_{c(b) j} \mathbbm{1}_{\{ b \in \Omega \}} + p_{ij}(t_b) \Bigl(V^{[1]}_{b_1 j} S_{b_2 j} + V^{[1]}_{b_2 j} S_{b_1 j}\Bigr)\biggr],
\label{eq17} \\[10pt]
V^{[2]}_{bi} = \sum_{j=1}^m \biggl[e^{[2]}_{ij}(h, t_b) F_{c(b) j} \mathbbm{1}_{\{ b \in \Omega \}} + p_{ij}(t_b) \Bigl(V^{[2]}_{b_1 j} S_{b_2 j} + V^{[2]}_{b_2 j} S_{b_1 j}\Bigr)\biggr],
\label{eq18} \\[10pt]
\begin{split}
W_{bi} = \sum_{j=1}^m \biggl[&2 \times e^{[1]}_{ij}(h, t_b) \Bigl(V^{[1]}_{b_1 j} S_{b_2 j} + V^{[1]}_{b_2 j} S_{b_1 j}\Bigr) \mathbbm{1}_{\{ b \in \Omega \}}\biggr. \\[-7.5pt]
&+ \biggl.p_{ij}(t_b) \Bigl(2 \times V^{[1]}_{b_1 j} V^{[1]}_{b_2 j} + W_{b_1 j} S_{b_2 j} + W_{b_2 j} S_{b_1 j}\Bigr)\biggr],
\end{split}
\label{eq19}
\end{gather}
respectively, where $b_1$ and $b_2$ represent the two branches that are ``below'' branch $b$ and $i = 1, ..., m$.
The recursive formulas presented in equations \eqref{eq17}-\eqref{eq19} can also be expressed as:
\begin{gather}
\mathbf{V}^{[1]}_b = \mathbf{e}^{[1]}(h, t_b) \mathbf{F}_{c(b)} \mathbbm{1}_{\{ b \in \Omega \}} + \mathbf{P}(t_b) \Bigl(\mathbf{V}^{[1]}_{b_1} \circ \mathbf{S}_{b_2} + \mathbf{V}^{[1]}_{b_2} \circ \mathbf{S}_{b_1}\Bigr),
\label{eq20} \\[10pt]
\mathbf{V}^{[2]}_b = \mathbf{e}^{[2]}(h, t_b) \mathbf{F}_{c(b)} \mathbbm{1}_{\{ b \in \Omega \}} + \mathbf{P}(t_b) \Bigl(\mathbf{V}^{[2]}_{b_1} \circ \mathbf{S}_{b_2} + \mathbf{V}^{[2]}_{b_2} \circ \mathbf{S}_{b_1}\Bigr),
\label{eq21} \\[10pt]
\begin{split}
\mathbf{W}_b = \ &2 \times \mathbf{e}^{[1]}(h, t_b) \Bigl(\mathbf{V}^{[1]}_{b_1} \circ \mathbf{S}_{b_2} + \mathbf{V}^{[1]}_{b_2} \circ \mathbf{S}_{b_1}\Bigr) \mathbbm{1}_{\{ b \in \Omega \}} \\
&+ \mathbf{P}(t_b) \Bigl(2 \times \mathbf{V}^{[1]}_{b_1} \circ \mathbf{V}^{[1]}_{b_2} + \mathbf{W}_{b_1} \circ \mathbf{S}_{b_2} + \mathbf{W}_{b_2} \circ \mathbf{S}_{b_1}\Bigr),
\end{split}
\label{eq22}
\end{gather}
respectively.
From \autoref*{vwlist}, we know that elements of $\mathbf{V}^{[1]}_b$ can be interpreted as the sum over all ``single-colored'' phylogenies for the subtree defined by $\Theta_b$ and for the predefined set of ``colored'' branches $\Omega_b = \Omega \cap \Theta_b$, conditional on the state of parent node $p(b)$.
Equation \eqref{eq20} partitions this sum into three distinct pieces: 1) $\mathbf{e}^{[1]}(h, t_b) \mathbf{F}_{c(b)} \mathbbm{1}_{\{ b \in \Omega \}}$, 2) $\mathbf{P}(t_b) \Bigl(\mathbf{V}^{[1]}_{b_1} \circ \mathbf{S}_{b_2}\Bigr)$, and 3) $\mathbf{P}(t_b) \Bigl(\mathbf{V}^{[1]}_{b_2} \circ \mathbf{S}_{b_1}\Bigr)$.
The first piece represents the ``single-colored'' tree with ``colored'' branch $b$ (only if $b \in \Omega$); the second piece represents a sum over ``single-colored'' phylogenies, where the ``colored'' branch is permuted across all branches in $\Omega_{b_1}$; and the third piece represents another sum over ``single-colored'' phylogenies, where the ``colored'' branch is permuted across all branches in $\Omega_{b_2}$.
This partitioning allows us to compute $\mathbf{V}^{[1]}_b$ as a function of previously cached vectors $\mathbf{F}_{c(b)}$, $\mathbf{S}_{b_1}$, $\mathbf{S}_{b_2}$, $\mathbf{V}^{[1]}_{b_1}$, and $\mathbf{V}^{[1]}_{b_2}$.
The recursions for $\mathbf{V}^{[2]}_b$ and $\mathbf{W}_b$ have analogous interpretations.
Note that our algorithm requires $O(n)$ storage because we cache a constant (with respect to $n$) number of vectors at all nodes $u \in \{ 1, ..., n-1, n, ..., 2n-1 \}$ and branches $b \in \Theta$.
In addition, our procedure utilizes $O(n)$ computations because there are $O(B_n) = O(n)$ iterations in the algorithm and each iteration involves a constant (with respect to $n$) number of operations.

\subsection{Posterior Mapping Variance Computation}

Our tips-to-root tree traversal procedure terminates after computing the vectors $\mathbf{F}_{root}$, $\mathbf{S}_{root_1}$, $\mathbf{S}_{root_2}$, $\mathbf{V}^{[1]}_{root_1}$, $\mathbf{V}^{[1]}_{root_2}$, $\mathbf{V}^{[2]}_{root_1}$, $\mathbf{V}^{[2]}_{root_2}$, $\mathbf{W}_{root_1}$, and $\mathbf{W}_{root_2}$, where $root$ denotes the root node label and $root_1$ and $root_2$ represent the two branches connecting the root node to its children.
We use these cached vectors to efficiently calculate the posterior mapping variance $\text{Var}(H_\Omega | \mathbf{D})$.
\par
We first describe how to compute the restricted mapping second moment $\text{E}(H_\Omega^2 \mathbbm{1}_{\mathbf{D}})$.
We remind our readers that the double sum over $b \neq b' \ (b,b' \in \Omega)$ in \eqref{eq9} can be visualized as a sum over all ``double-colored'' phylogenies, where $\Omega$ denotes the predefined set of ``colored'' branches.
There are four distinct cases of $b \neq b'$ that we consider in \eqref{eq9}: 1) $b \preceq root_1, b' \preceq root_1$; 2) $b \preceq root_1, b' \preceq root_2$; 3) $b \preceq root_2, b' \preceq root_1$; and 4) $b \preceq root_2, b' \preceq root_2$.
Cases 1) and 4) represent the ``double-colored'' phylogenies that have both ``colored'' branches on the same side of the root, while Cases 2) and 3) denote the ``double-colored'' phylogenies that have one ``colored'' branch on each side of the root.
This decomposition of the double sum in \eqref{eq9} was suggested by \citet{kenney2012hessian} in the context of computing second derivatives of phylogenetic likelihood functions.
The sums over all ``double-colored'' phylogenies in Cases 1), 2), 3), and 4) are mathematically represented as $\boldsymbol{\pi}^T \Bigl(\mathbf{W}_{root_1} \circ \mathbf{S}_{root_2}\Bigr)$, $\boldsymbol{\pi}^T \Bigl(\mathbf{V}^{[1]}_{root_1} \circ \mathbf{V}^{[1]}_{root_2}\Bigr)$, $\boldsymbol{\pi}^T \Bigl(\mathbf{V}^{[1]}_{root_1} \circ \mathbf{V}^{[1]}_{root_2}\Bigr)$, and $\boldsymbol{\pi}^T \Bigl(\mathbf{W}_{root_2} \circ \mathbf{S}_{root_1}\Bigr)$, respectively.
Thus, the double sum in \eqref{eq9} can be efficiently computed as:
\begin{equation}
\boldsymbol{\pi}^T \Bigl(2 \times \mathbf{V}^{[1]}_{root_1} \circ \mathbf{V}^{[1]}_{root_2} + \mathbf{W}_{root_1} \circ \mathbf{S}_{root_2} + \mathbf{W}_{root_2} \circ \mathbf{S}_{root_1}\Bigr).
\label{eq23}
\end{equation}
The sum over $b \in \Omega$ in \eqref{eq8} can also be calculated using the cached vectors mentioned above.
We can visualize the sum in \eqref{eq8} as a sum over all ``single-colored'' phylogenies, where $\Omega$ denotes the predefined set of ``colored'' branches.
There are two cases of $b \in \Omega$ that we consider in \eqref{eq8}: 1) $b \preceq root_1$ and 2) $b \preceq root_2$.
Using logic similar to that described above, we can calculate the sum in \eqref{eq8} as:
\begin{equation}
\boldsymbol{\pi}^T \Bigl[\bigl(\mathbf{V}^{[1]}_{root_1} + \mathbf{V}^{[2]}_{root_1}\bigr) \circ \mathbf{S}_{root_2} + \bigl(\mathbf{V}^{[1]}_{root_2} + \mathbf{V}^{[2]}_{root_2}\bigr) \circ \mathbf{S}_{root_1}\Bigr].
\label{eq24}
\end{equation}
We obtain a simple formula for computing the restricted mapping second moment $\text{E}(H_\Omega^2 \mathbbm{1}_{\mathbf{D}})$ by adding together the expressions in \eqref{eq23} and \eqref{eq24}:
\begin{equation}
\begin{split}
\text{E}(H_\Omega^2 \mathbbm{1}_{\mathbf{D}}) = \boldsymbol{\pi}^T \Bigl[&2 \times \mathbf{V}^{[1]}_{root_1} \circ \mathbf{V}^{[1]}_{root_2} + \mathbf{W}_{root_1} \circ \mathbf{S}_{root_2} + \mathbf{W}_{root_2} \circ \mathbf{S}_{root_1}\Bigr. \\
&+ \Bigl.\bigl(\mathbf{V}^{[1]}_{root_1} + \mathbf{V}^{[2]}_{root_1}\bigr) \circ \mathbf{S}_{root_2} + \bigl(\mathbf{V}^{[1]}_{root_2} + \mathbf{V}^{[2]}_{root_2}\bigr) \circ \mathbf{S}_{root_1}\Bigr].
\end{split}
\label{eq25}
\end{equation}
\par
Other quantities involved in the calculation of $\text{Var}(H_\Omega | \mathbf{D})$ include the phylogenetic likelihood $\text{P}(\mathbf{D})$ and the restricted mapping first moment $\text{E}(H_\Omega \mathbbm{1}_{\mathbf{D}})$.
From \citep{felsenstein1981evolutionary}, we know that $\text{P}(\mathbf{D}) = \boldsymbol{\pi}^T \mathbf{F}_{root}$.
\citet{minin2008fast} express $\text{E}(H_\Omega \mathbbm{1}_{\mathbf{D}})$ in the following manner:
\begin{equation}
\text{E}(H_\Omega \mathbbm{1}_{\mathbf{D}}) = \sum_b \sum_{\mathbf{i}} e^{[1]}_{i^*_{p(b)} i^*_{c(b)}}(h, t_b) \pi_{i^*_1} \prod_{b^* \in \Theta \setminus \{ b \}} p_{i^*_{p(b^*)} i^*_{c(b^*)}}(t_{b^*}),
\label{eq26}
\end{equation}
where $b \in \Omega$.
Notice that the right-hand side of equation \eqref{eq26} is virtually identical to the sum over $b \in \Omega$ in \eqref{eq8}.
Given our interpretations of the sums in \eqref{eq8} and \eqref{eq9}, it is easy to see that:
\begin{equation}
\text{E}(H_\Omega \mathbbm{1}_{\mathbf{D}}) = \boldsymbol{\pi}^T \Bigl(\mathbf{V}^{[1]}_{root_1} \circ \mathbf{S}_{root_2} + \mathbf{V}^{[1]}_{root_2} \circ \mathbf{S}_{root_1}\Bigr).
\label{eq27}
\end{equation}
\par
Finally, the posterior mapping variance $\text{Var}(H_\Omega | \mathbf{D})$ can be expressed as follows:
\begin{align}
\text{Var}(H_\Omega | \mathbf{D}) &= \text{E}(H_\Omega^2 | \mathbf{D}) - \text{E}(H_\Omega | \mathbf{D})^2
\label{eq28} \\
&= \frac{\text{E}(H_\Omega^2 \mathbbm{1}_{\mathbf{D}})}{\text{P}(\mathbf{D})} - \biggl[\frac{\text{E}(H_\Omega \mathbbm{1}_{\mathbf{D}})}{\text{P}(\mathbf{D})}\biggr]^2.
\label{eq29}
\end{align}
We can compute $\text{Var}(H_\Omega | \mathbf{D})$ by first calculating $\text{P}(\mathbf{D})$, $\text{E}(H_\Omega \mathbbm{1}_{\mathbf{D}})$, and $\text{E}(H_\Omega^2 \mathbbm{1}_{\mathbf{D}})$ using our post-order tree traversal algorithm and then substituting these quantities into equation \eqref{eq29}.

\subsection{Prior Mapping Variance Computation}

Given our efficient calculation of the posterior mapping variance $\text{Var}(H_\Omega | \mathbf{D})$, it is natural to ask whether it is possible to extend the tree traversal algorithm described above to the computation of the prior mapping variance $\text{Var}(H_\Omega)$.
There is a plethora of literature in evolutionary biology that discusses how to calculate the prior mapping variance for a single tree branch \citep{zheng2001dispersion, bloom2007thermodynamics, minin2008counting, minin2008fast}.
\citet{siepel2006new} describe a dynamic programming procedure that approximates the probability distribution of $H_\Theta$ and use it to estimate $\text{Var}(H_\Theta)$.
\par
It turns out that we can exactly compute prior variances using a modified version of our tree traversal algorithm.
The only changes that need to be made are to the initializations of $\mathbf{F}_u$ at all terminal nodes $u$.
If we set $F_{ui} = 1$ for all terminal nodes $u$ and $i = 1, ..., m$, then our recursive procedure will be able to compute the prior variance $\text{Var}(H_\Omega)$.
We also note that all elements of the vectors $\mathbf{F}_u$ and $\mathbf{S}_b$ for $u \in \{ 1, ..., n-1, n, ..., 2n-1 \}$ and $b \in \Theta$ are equal to 1 as a result of these modified initializations.
Remember that the initializations of $\mathbf{F}_u$ can account for partially observed and/or missing data at the tips of $\tau$.
These modified $\mathbf{F}_u$ initializations are intuitive because prior mapping moments average over all possible observed trait values at the tips of the phylogeny and thus every combination of observed tip states needs to be accounted for in our $\mathbf{F}_u$ initializations.
As we will see, the prior mapping variance $\text{Var}(H_\Omega)$ is essential to one of the applications we present in this paper.

\subsection{Prior and Posterior Mapping Covariances}

Two other quantities of interest associated with stochastic mapping summaries are the prior mapping covariance $\text{Cov}(H_{\Omega_1}, H_{\Omega_2})$ and the posterior mapping covariance $\text{Cov}(H_{\Omega_1}, H_{\Omega_2} | \mathbf{D})$, where $\Omega_1, \Omega_2 \subseteq \Theta$ are predefined sets of branch indices.
Efficient computations of the prior and posterior mapping covariances are interesting in their own right and utilized in one of the applications we present in this paper.
In the Appendix, we describe another tree traversal algorithm for computing these covariances.
This new algorithm is a generalization of the recursive procedure described above and much of the intuition provided for our original procedure carries over to this new algorithm.
Furthermore, this new procedure runs in $O(n)$ time and with $O(n)$ storage.

\subsection{Implementation}

The efficient calculations of the prior and posterior mapping moments discussed above are implemented in the R package \verb|phylomoments|, which is available at \url{https://github.com/dunleavy005/phylomoments}.
This package also contains our implementation of the stochastic mapping simulation technique put forth by \citet{nielsen2002mapping} along with other assorted functions.
The computationally intensive parts of our methods are written in C++ and ported to R using the R packages \verb|Rcpp| and \verb|RcppArmadillo| \citep{eddelbuettel2011rcpp, eddelbuettel2014rcpparmadillo}.
In the next section, we present two scientific applications that employ stochastic mapping variances.

\section{Applications}

\subsection{Testing for Rate Variation Across Sites}

Our first application is centered around an across-site rate variation test proposed by \citet{nielsen2002mapping}.
\citet{nielsen2002mapping} uses simulated posterior mapping variances within a posterior predictive approach to model diagnostics.
In this subsection, we describe a posterior predictive testing framework that adheres to the principles outlined by \citet{gelman1996posterior} and test for across-site rate variation in two real datasets using exactly computed posterior mapping variances.

\subsubsection{Overview of Posterior Predictive Tests}

Conceptually, posterior predictive assessments can be seen as Bayesian analogues of classical frequentist model diagnostics and hypothesis tests.
Unlike classical testing procedures, posterior predictive tests permit the use of ``test statistics'' that depend on both data and unknown parameters.
These ``test statistics'' are otherwise known as discrepancy measures \citep{meng1994posterior, gelman1996posterior}.
In this subsection, we denote the discrepancy measure as $T \equiv T(\mathbf{D}_{1:L}, \boldsymbol{\theta})$.
Posterior predictive model testing is based on the following principle: if the assumed model adequately fits the observed data $\mathbf{D}^{obs}_{1:L}$, then simulated datasets $\mathbf{D}_{1:L}^{rep}$ from the assumed model should look like $\mathbf{D}_{1:L}^{obs}$.
Similarity between $\mathbf{D}_{1:L}^{obs}$ and $\mathbf{D}_{1:L}^{rep}$ is measured through the discrepancy $T$.
We would like to compare the observed discrepancy $T(\mathbf{D}^{obs}_{1:L}, \boldsymbol{\theta})$ to a reference distribution induced by the hypothesized model.
The reference distribution used in posterior predictive diagnostics is derived from the joint posterior distribution of $\mathbf{D}_{1:L}^{rep}$ and $\boldsymbol{\theta}$:
\begin{equation}
\text{P}(\mathbf{D}_{1:L}^{rep}, \boldsymbol{\theta} | \mathbf{D}_{1:L}^{obs}) = \text{P}(\mathbf{D}_{1:L}^{rep} | \boldsymbol{\theta}) \text{P}(\boldsymbol{\theta} | \mathbf{D}_{1:L}^{obs}).
\label{eq30}
\end{equation}
Intuitively, this distribution indicates which datasets and parameter values are most plausible if the assumed model holds true.
Posterior predictive assessments are usually conducted using simulations as the joint distribution in \eqref{eq30} is often analytically intractable.
We summarize posterior predictive simulations in the following three steps:
\begin{enumerate}
\item Sample $\boldsymbol{\theta}^* \sim \text{P}(\boldsymbol{\theta} | \mathbf{D}_{1:L}^{obs})$;
\item Simulate $\mathbf{D}_{1:L}^{rep,*} \sim \text{P}(\mathbf{D}_{1:L}^{rep} | \boldsymbol{\theta}^*)$;
\item Calculate $T(\mathbf{D}^{obs}_{1:L}, \boldsymbol{\theta}^*)$ and $T(\mathbf{D}_{1:L}^{rep,*}, \boldsymbol{\theta}^*)$.
\end{enumerate}
We repeat these steps $N$ times, where $N$ is a large number, and then compare the $N$ samples of $T(\mathbf{D}^{obs}_{1:L}, \boldsymbol{\theta}^*)$ and $T(\mathbf{D}_{1:L}^{rep,*}, \boldsymbol{\theta}^*)$ by constructing two empirical distributions; a small overlap between these distributions, which could be visualized with histograms, suggests a poor model fit.
We can quantify the disagreement between the observed and predicted discrepancies by calculating the posterior predictive $p$-value \citep{meng1994posterior, gelman1996posterior}:
\begin{equation}
ppp = \text{P}\bigl[T(\mathbf{D}_{1:L}^{rep}, \boldsymbol{\theta}) > T(\mathbf{D}^{obs}_{1:L}, \boldsymbol{\theta}) {\big |} \mathbf{D}^{obs}_{1:L}\bigr],
\label{eq31}
\end{equation}
where the probability is computed with respect to the joint posterior distribution $\text{P}(\mathbf{D}_{1:L}^{rep}, \boldsymbol{\theta} | \mathbf{D}_{1:L}^{obs})$.
Given $N$ posterior samples of $\mathbf{D}_{1:L}^{rep}$ and $\boldsymbol{\theta}$, we can estimate the posterior predictive $p$-value by computing:
\begin{equation}
ppp \approx \frac{1}{N} \sum_{i=1}^N \mathbbm{1}_{\bigl\{ T(\mathbf{D}_{1:L}^{rep,(i)}, \boldsymbol{\theta}^{(i)}) > T(\mathbf{D}^{obs}_{1:L}, \boldsymbol{\theta}^{(i)}) \bigr\}},
\label{eq32}
\end{equation}
where $\mathbf{D}_{1:L}^{rep,(i)}$ and $\boldsymbol{\theta}^{(i)}$ represent the $i$th posterior samples of $\mathbf{D}_{1:L}^{rep}$ and $\boldsymbol{\theta}$, respectively, for $i = 1, ..., N$.

\subsubsection{Posterior Predictive Rate Variation Tests}

Now, we use the posterior predictive testing framework described above to formulate a test for rate variation across sites in an alignment.
We assume that alignment sites evolve independently according to the same distribution
provided by the continuous-time Markov process $\psi_{\boldsymbol{\theta}}$.
We focus on selecting discrepancy measures that can gauge the variability in substitution rates across sites.
One possible discrepancy measure is the variance of the total number of substitutions in the alignment \citep{nielsen2002mapping}:
\begin{equation}
T_{var} \equiv T_{var}(\mathbf{D}_{1:L}, \boldsymbol{\theta}) = \text{Var}\Biggl(\sum_{i=1}^L H^{(i)}_\Theta {\Bigg |} \mathbf{D}_{1:L}\Biggr) = \sum_{i=1}^L \text{Var}(H_\Theta | \mathbf{D}_i),
\label{eq33}
\end{equation}
where $H^{(i)}_\Theta$ represents the number of substitutions at site $i$.
Note that the second equality in \eqref{eq33} follows from conditional independence assumptions.
Another discrepancy we consider in our analyses is the posterior dispersion index (i.e.\ posterior variance-to-mean ratio) for substitution counts:
\begin{align}
\begin{split}
T_{disp} \equiv T_{disp}(\mathbf{D}_{1:L}, \boldsymbol{\theta}) &= \text{Var}\Biggl(\sum_{i=1}^L H^{(i)}_\Theta {\Bigg |} \mathbf{D}_{1:L}\Biggr) {\Bigg /} \text{E}\Biggl(\sum_{i=1}^L H^{(i)}_\Theta {\Bigg |} \mathbf{D}_{1:L}\Biggr) \\[5pt]
&= \Biggl[\sum_{i=1}^L \text{Var}(H_\Theta | \mathbf{D}_i)\Biggr] {\Bigg /} \Biggl[\sum_{i=1}^L \text{E}(H_\Theta | \mathbf{D}_i)\Biggr],
\end{split}
\label{eq34}
\end{align}
where $H^{(i)}_\Theta$ is specified as above.
In the presence of rate variation across sites, the prior dispersion index for substitution counts is often greater than 1 \citep{yang1996among}.
One way to see this is by assuming that the number of substitutions occurring on a particular tree branch follows a Poisson distribution, where the Poisson rate parameter varies across sites according to a gamma distribution.
A simple calculation shows that, marginally, the number of substitutions occurring on this branch follows a negative binomial distribution, which has a variance-to-mean ratio that is greater than 1.
We can make a similar argument about the number of substitutions across an entire phylogeny.
Even though the above explanation applies only to the prior dispersion index, it does motivate our use of the posterior dispersion index as a discrepancy measure for detecting across-site rate variation in an alignment.
\par
The posterior predictive simulations for our rate variation test can be summarized using the same three steps described above.
We can sample $\boldsymbol{\theta}^* \sim \text{P}(\boldsymbol{\theta} | \mathbf{D}_{1:L}^{obs})$ using a Bayesian phylogenetic inference software package; in our examples, we use the computer program MrBayes \citep{huelsenbeck2001mrbayes} to perform the posterior sampling of $\boldsymbol{\theta}$.
For every posterior sample of $\boldsymbol{\theta}$, we can simulate a replicate alignment $\mathbf{D}_{1:L}^{rep,*} \sim \text{P}(\mathbf{D}_{1:L}^{rep} | \boldsymbol{\theta}^*)$ by independently generating $\mathbf{D}^{rep,*}_1, ..., \mathbf{D}^{rep,*}_L$ according to $\psi_{\boldsymbol{\theta}^*}$; we simulate tip data from $\psi_{\boldsymbol{\theta}^*}$ using the standard discretized CTMC approach \citep[Chapter 9]{yang2006computational}.
We can then analytically calculate the observed and predicted discrepancies (i.e.\ $T(\mathbf{D}^{obs}_{1:L}, \boldsymbol{\theta}^*)$ and $T(\mathbf{D}_{1:L}^{rep,*}, \boldsymbol{\theta}^*)$) for $T = T_{var}, T_{disp}$ using the algorithm presented in the previous section.
\citet{nielsen2002mapping} also used the discrepancy $T_{var}$ in posterior predictive rate variation tests but could only obtain Monte Carlo estimates of $T_{var}$.
Monte Carlo estimation of $T_{var}$ is computationally intensive because it uses simulations to estimate $\text{Var}(H_\Theta | \mathbf{D}_i)$ for $i = 1, ..., L$.
The tree traversal algorithm proposed in this paper not only eliminates the Monte Carlo error associated with estimating the variance $T_{var}$ but also speeds up the computation of this quantity.

\subsubsection{Testing for Rate Variation in Two Sequence Alignments}

We analyze the two sequence alignments used by \citet{nielsen2002mapping} to demonstrate the effectiveness of our posterior predictive testing scheme.
Our first dataset contains $\beta$-globin sequences for 17 vertebrate species, where each sequence is 432 base pairs long.
Our second dataset comprises 28 sequences of the hemagglutinin (HA) gene of human influenza virus A; each sequence has 987 base pairs.
In all our analyses, we use a general time-reversible (GTR) substitution model \citep{tavare1986some} with a Dirichlet(1,1,1,1,1,1) prior for the GTR exchangeability rates and a Dirichlet(1,1,1,1) prior for the base frequencies $\boldsymbol{\pi}$.
Furthermore, we assume a uniform prior on all possible tree topologies $\tau$ and let all branch lengths in $\mathbf{t}$ be a priori uniformly distributed on the interval $[0,100]$.
For each dataset, we generate $N = 1000$ posterior samples of $\mathbf{D}_{1:L}^{rep}$ and $\boldsymbol{\theta}$ using the simulation procedure described previously and calculate the observed and predicted values of $T_{var}$ and $T_{disp}$.
The 1000 posterior samples of $\boldsymbol{\theta}$ are obtained by running the Markov chain Monte Carlo (MCMC) procedure in MrBayes for 1,100,000 iterations and storing values of $\boldsymbol{\theta}$ every 1000 iterations from iteration 101,000 to iteration 1,100,000.
We use trace plots (not shown in this paper) to assess convergence of the MCMC samplers and find that using 1,100,000 MCMC iterations is sufficient for our purposes.
\begin{figure}[!ht]
\centering
\includegraphics[width=\textwidth, trim=0 10 0 40, clip]{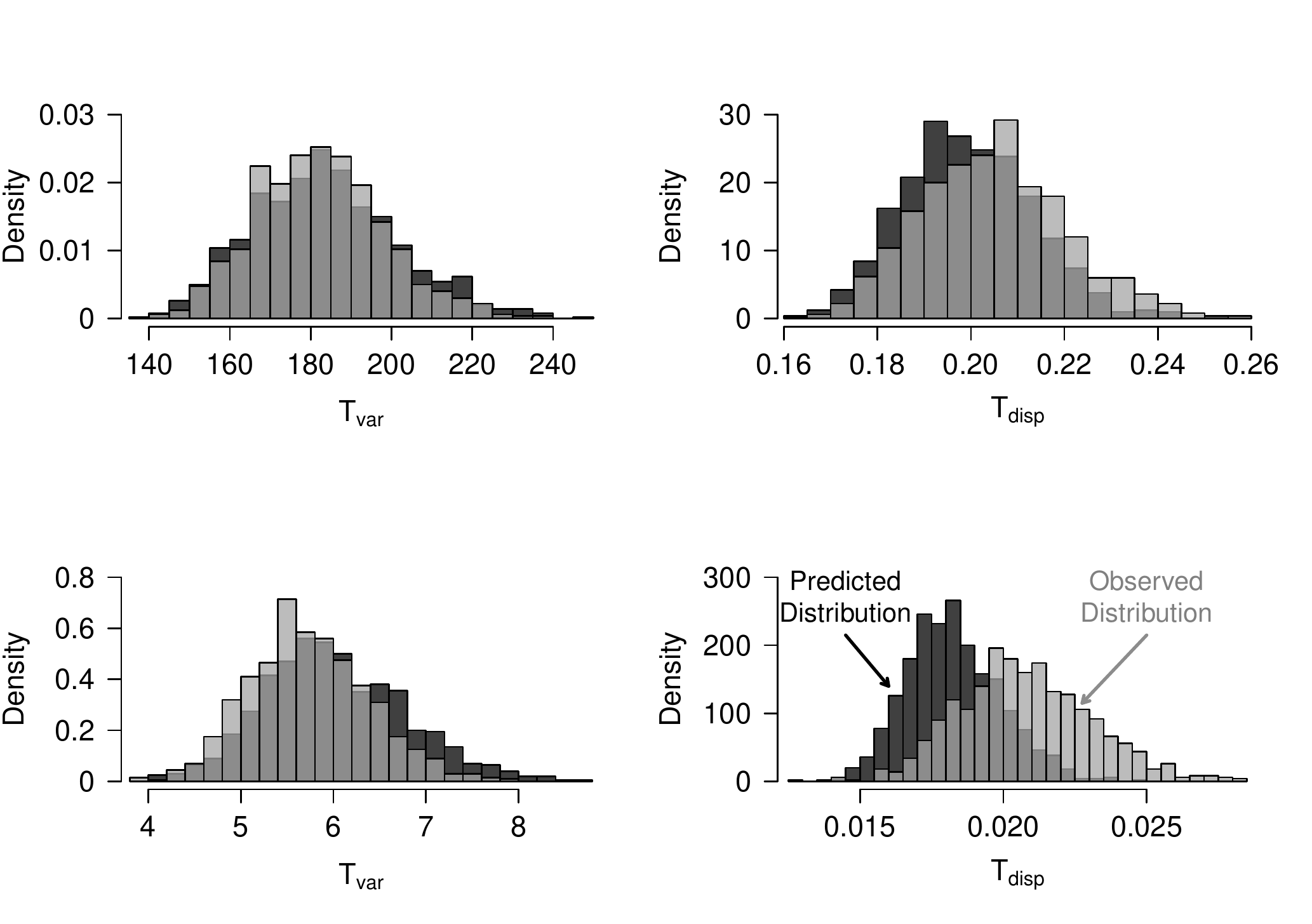}
\caption{Observed and predicted distributions of the discrepancies $T_{var}$ and $T_{disp}$.  In each plot, we superimpose the observed distribution (grey) on top of the predicted distribution (black).  The top two plots display the observed and predicted distributions for the $\beta$-globin dataset, while the bottom two plots show the observed and predicted distributions for the influenza dataset.  These distributions were constructed using $N = 1000$ posterior samples of $\mathbf{D}_{1:L}^{rep}$ and $\boldsymbol{\theta}$.}
\label{obspreddist}
\end{figure}
\par
We summarize the observed and predicted discrepancies in \autoref*{obspreddist}, which 
demonstrates that the observed distributions of $T_{var}$ do not deviate much from the corresponding predicted distributions of $T_{var}$.
The posterior predictive $p$-values that were computed using the discrepancy $T_{var}$ are approximately 0.60 and 0.80 for the $\beta$-globin and influenza datasets, respectively.
Thus, it seems that our posterior predictive test based on $T_{var}$ fails to detect across-site rate variation in the two datasets.
We note that our posterior predictive test results based on $T_{var}$ do not corroborate the findings by \citet{nielsen2002mapping}, who concluded that there is rate variation across sites in the $\beta$-globin and influenza datasets.
However, \citet{nielsen2002mapping} used the posterior variance of substitution counts in a slightly different way without clearly specifying all predictive distributions.
We believe our analyses are better aligned with the posterior predictive principles outlined by \citet{gelman1996posterior}.
\par
In \autoref*{obspreddist}, the observed and predicted distributions of $T_{disp}$ do not completely overlap and appear more separated than the observed and predicted distributions of $T_{var}$.
For both datasets, the observed values of $T_{disp}$ are, on average, greater than the predicted values of $T_{disp}$.
This suggests that the discrepancy $T_{disp}$ is able to detect observed rate variation that is not accounted for by our hypothesized model.
The posterior predictive $p$-values that were computed using $T_{disp}$ are approximately 0.17 and 0.031 for the $\beta$-globin and influenza datasets, respectively; note that these $p$-values are smaller than the corresponding $p$-values that were computed using $T_{var}$ and as a result provide stronger evidence in support of the rate variation hypothesis.
\par
A popular frequentist approach to modeling rate variation among sites employs a discrete gamma distribution with a fixed number of rate classes \citep{yang1994maximum, yang1996among}.
We check our posterior predictive test results by performing likelihood ratio tests that compare discrete gamma models with one rate category ($H_0$) and four rate categories ($H_a$).
We perform the likelihood ratio tests using the PhyML package \citep{guindon2010new} and obtain $p$-values close to 0; these tests also suggest the presence of across-site rate variation in the $\beta$-globin and influenza datasets.
The posterior predictive $p$-values that were computed using $T_{disp}$ are not as small as the likelihood ratio $p$-values, but this should not be surprising because posterior predictive $p$-values tend to be more conservative than classical frequentist $p$-values \citep{meng1994posterior}.
Our posterior predictive analyses suggest that the posterior dispersion index $T_{disp}$ is better than the posterior variance $T_{var}$ at detecting observed rate variation among sites.

\subsubsection{Monte Carlo Error and Timing Analyses}

To assess the efficiency gains from using our tree traversal algorithm in this setting, we approximate Monte Carlo standard errors and running times associated with simulation-based $T_{var}$ estimates.
For each dataset, we compute these standard errors and running times on randomly subsampled alignments of length $L$ using $m$ Monte Carlo replicates per site, where $L = 50, 100, 200, 400$ and $m = 100, 500, 1000, 10000$; the Monte Carlo standard errors are approximated using well-known formulas for the moments of the sample variance \citep[Chapter VI]{mood1974introduction}.
We account for the posterior uncertainty in $\boldsymbol{\theta}$ by first calculating Monte Carlo errors and running times for 200 randomly subsampled posterior $\boldsymbol{\theta}$'s and then averaging these metrics across the $\boldsymbol{\theta}$ samples; the same 200 samples of $\boldsymbol{\theta}$ are used in all our simulations.
For each setting of $L$, we also compute the exact values of $T_{var}$ and track the corresponding running times for the 200 posterior samples of $\boldsymbol{\theta}$ and then average these results over the subsampled $\boldsymbol{\theta}$'s.
\par
Tables \ref{mctable} and \ref{mctablesupp} present the running time comparisons and Monte Carlo error approximations, respectively, for the $\beta$-globin and influenza datasets; \autoref*{mctablesupp} can be found in the Appendix.
Based on the results shown in \autoref*{mctablesupp}, it seems that the Monte Carlo error has a convergence rate of $O\Bigl(\sqrt{L/m}\Bigr)$.
This can be justified using a Central Limit Theorem argument if $L$ and $m$ are large; remember that the Monte Carlo estimator of $T_{var}$ is a sum of $L$ independent, non-identically distributed sample variances, where each sample variance is calculated using $m$ independent Monte Carlo replicates.
Furthermore, the Monte Carlo error percentages range from 0.15\% to 28\% across the two datasets.
\autoref{mctable} suggests that the Monte Carlo running time increases linearly in $L$ and $m$ as we might expect.
We can also see that our exact computations of $T_{var}$ are at least an order of magnitude faster than the Monte Carlo estimates of $T_{var}$.
Thus, it is clear that our tree traversal algorithm improves the computational efficiency of posterior predictive rate variation tests that utilize Monte Carlo discrepancy estimates.
\begin{table}[!ht]
\centering
\begin{tabular}{cccccccccc}
\hline
& \multicolumn{4}{c}{Running times ($\beta$-globin)} & & \multicolumn{4}{c}{Running times (influenza)} \\
\cline{2-5} \cline{7-10}
& $L = 50$ & $L = 100$ & $L = 200$ & $L = 400$ & & $L = 50$ & $L = 100$ & $L = 200$ & $L = 400$ \\
\hline
$m = 100$ & 0.70 & 1.4 & 3.1 & 6.1 & & 1.0 & 2.0 & 4.1 & 8.0 \\
$m = 500$ & 3.3 & 6.5 & 15 & 28 & & 4.7 & 9.3 & 19 & 37 \\
$m = 1000$ & 7.2 & 13 & 28 & 57 & & 9.0 & 18 & 37 & 74 \\
$m = 10000$ & 72 & 140 & 290 & 570 & & 93 & 180 & 370 & 740 \\
Exact & 0.0040 & 0.0059 & 0.0097 & 0.017 & & 0.0061 & 0.0091 & 0.015 & 0.026 \\
\hline
\end{tabular}
\caption{Monte Carlo running times associated with simulation-based $T_{var}$ estimates for the $\beta$-globin and influenza datasets.  We compute these running times on randomly subsampled alignments of length $L$ using $m$ Monte Carlo replicates per site.  Each table entry, excluding the entries on the bottom row, represents an averaged Monte Carlo running time (in seconds), where the averaging is done over 200 randomly subsampled posterior $\boldsymbol{\theta}$'s.  Each table entry on the bottom row denotes an average over running times (in seconds) associated with exact calculations of $T_{var}$, where the averaging is done over the same 200 posterior samples of $\boldsymbol{\theta}$ mentioned previously.  All table entries are rounded to two significant digits.}
\label{mctable}
\end{table}
\vspace{-3.75mm}

\subsection{Detecting Evolutionarily Conserved Regions in Genomic Alignments}

Our second application focuses on the detection of conserved elements in genomic alignments; in this setting, conservation refers to evolution that is slower than expected.
Detection of conserved DNA sites is of prime interest in comparative genomics because most of the conserved elements in genome-wide sequence alignments are believed to be caused by negative selection and to have evolutionarily important biological functions \citep{siepel2005evolutionarily}.
Computational methods for detecting conserved genomic regions are essential because they are used to flag candidate functional elements, which can then be further examined experimentally \citep{siepel2006new}.
\par
In this subsection, we analyze two statistical tests of conservation presented by \citet{siepel2006new}.
One test is used to detect conservation across all lineages in a phylogeny, while the other is used to identify lineage-specific conservation; both tests are referred to as SPH conservation tests and implemented in the computer program phyloP \citep{pollard2010detection}.
Our exact calculations of prior and posterior mapping moments can be used to make the SPH conservation tests more powerful.
We present some modifications to these conservation tests and demonstrate the efficacy of our proposed changes via simulations.

\subsubsection{Modifying the SPH Conservation Tests}

Let $\psi_{\boldsymbol{\theta}_{\text{neut}}}$ denote the baseline neutral evolutionary model that is assumed to be given; neutral models are commonly estimated using fourfold degenerate sites extracted from genome-wide sequence alignments of interest \citep{pollard2010detection}.
As in \citep{pollard2010detection}, we define $\psi(\rho, \lambda; \Theta_b)$ to be a scaled evolutionary model that is identical to $\psi_{\boldsymbol{\theta}_{\text{neut}}}$ except that it has all its branch lengths scaled by the factor $\rho \in [0,1]$ and the branch lengths in the subtree defined by $\Theta_b$ additionally scaled by the factor $\lambda \in [0,1]$.
For convenience, we let $\psi(\rho) \equiv \psi(\rho, \lambda = 1; \Theta_b)$ for all $b \in \Theta$ and $\rho \in [0,1]$.
Throughout this subsection, we assume that $\mathbf{D}_1, ..., \mathbf{D}_L$ are independent and identically distributed according to $\psi(\cdot)$.
The two tests of conservation described in \citep{siepel2006new} reduce to tests of the models that constrain parameters $\rho$ and $\lambda$ to particular values.
\par
The SPH ``all-branch'' conservation test examines conservation across all branches of the phylogeny.
Specifically, it tests the null hypothesis $H_0: \rho = 1$ against the alternative hypothesis $H_a: \rho < 1$ for the evolutionary model $\psi(\rho)$.
\citet{siepel2006new} use the following test statistic to distinguish between the two hypotheses:
\begin{equation}
\mathcal{T}_{all} \equiv \mathcal{T}_{all}(\mathbf{D}_{1:L}) = \text{E}\Biggl(\sum_{i=1}^L H^{(i)}_\Theta {\Bigg |} \mathbf{D}_{1:L}\Biggr) = \sum_{i=1}^L \text{E}(H_\Theta | \mathbf{D}_i),
\label{eq35}
\end{equation}
where $H^{(i)}_\Theta$ represents the number of substitutions at site $i$.
The test statistic in \eqref{eq35} serves as a proxy for the ``observed'' number of substitutions in the genomic alignment $\mathbf{D}_{1:L}$.
The prior distribution of $\sum_{i=1}^L H^{(i)}_\Theta$ is taken to be the null distribution, and the $p$-value of this test is obtained by first comparing the observed value of $\mathcal{T}_{all}$ to this prior distribution and then computing the corresponding left-tail probability.
It turns out that the $p$-values obtained from this testing procedure are not uniformly distributed and tend to be conservative under the null hypothesis \citep{siepel2006new}.
\par
To understand why this occurs, we must examine the null distribution chosen for this test.
Notice first that $\mathcal{T}_{all}$ is a sum of independent and identically distributed random variables with mean $\text{E}[\text{E}(H_\Theta | \mathbf{D})] = \text{E}(H_\Theta)$ and variance $\text{Var}[\text{E}(H_\Theta | \mathbf{D})] = \text{Var}(H_\Theta) - \text{E}[\text{Var}(H_\Theta | \mathbf{D})]$.
For $L$ large, we can invoke the Central Limit Theorem and approximate the sampling distribution of $\mathcal{T}_{all}$ with a normal distribution having mean $L \times \text{E}(H_\Theta)$ and variance $L \times \text{Var}(H_\Theta) - L \times \text{E}[\text{Var}(H_\Theta | \mathbf{D})]$.
Using a similar asymptotic argument as above, we can also approximate the null distribution of $\sum_{i=1}^L H^{(i)}_\Theta$ --- an unobserved quantity --- with a normal distribution having mean $L \times \text{E}(H_\Theta)$ and variance $L \times \text{Var}(H_\Theta)$.
\citet{siepel2006new} use the exact version of this distribution, which results in the conservative nature of their all-branch test, because the correct null distribution of $\mathcal{T}_{all}$, at least asymptotically, has a variance that is smaller than the one assumed by the authors.
For a given significance level, an overdispersed null distribution causes the critical value for rejecting $H_0$ to be more extreme than it would be for a proper null distribution.
This is of practical importance because more extreme critical values make it harder to correctly flag conserved genomic elements.
\par
The SPH all-branch conservation test can be corrected and made more powerful by using the correct asymptotic distribution of $\mathcal{T}_{all}$ as the null distribution.
The dynamic programming algorithm discussed in this paper can be used to calculate the mean and variance of this asymptotic distribution.
The quantities $\text{E}(H_\Theta)$ and $\text{Var}(H_\Theta)$ are easily computed using our results for prior mapping moments.
Given our efficient computation of the posterior mapping variance $\text{Var}(H_\Theta | \mathbf{D})$, we estimate $\text{E}[\text{Var}(H_\Theta | \mathbf{D})]$ using Monte Carlo simulation of sequence data $\mathbf{D}$.
Even though we need Monte Carlo simulations to calculate the asymptotic variance of $\mathcal{T}_{all}$, our approach is still more efficient than directly estimating $\text{Var}[\text{E}(H_\Theta | \mathbf{D})]$ via Monte Carlo sampling.
\par
\citet{siepel2006new} also describe two testing procedures that are used to detect lineage-specific conservation.
Both procedures analyze conservation at the subtree level and are referred to as SPH ``subtree'' tests.
Formally, these two approaches test the null hypothesis $H_0: \lambda = 1, \rho \in [0,1]$ against the alternative hypothesis $H_a: \lambda < 1, \rho \in [0,1]$ for the evolutionary model $\psi(\rho, \lambda; \Theta_b)$.
The following test statistic is used in both procedures:
\begin{equation}
\mathcal{T}_{sub} \equiv \mathcal{T}_{sub}(\mathbf{D}_{1:L}) = \text{E}\Biggl(\sum_{i=1}^L H^{(i)}_{\Theta_b} {\Bigg |} \mathbf{D}_{1:L}\Biggr) = \sum_{i=1}^L \text{E}(H_{\Theta_b} | \mathbf{D}_i),
\label{eq36}
\end{equation}
where $H^{(i)}_{\Theta_b}$ denotes the number of substitutions in the subtree defined by $\Theta_b$ at site $i$.
Note that the statistic in \eqref{eq36} is the subtree equivalent of that shown in \eqref{eq35}.
The SPH ``marginal'' subtree test compares the observed value of $\mathcal{T}_{sub}$ to the marginal distribution of $\sum_{i=1}^L H^{(i)}_{\Theta_b}$, while the SPH ``conditional'' subtree test compares the observed value of $\mathcal{T}_{sub}$ to the conditional distribution of $\sum_{i=1}^L H^{(i)}_{\Theta_b}$ given $\sum_{i=1}^L H^{(i)}_\Theta$ is equal to the observed value of $\mathcal{T}_{all}$.
Conceptually, the marginal subtree test examines whether the number of substitutions in the subtree is less than would be expected under the null model, whereas the conditional subtree test analyzes whether the number of substitutions in the subtree is surprising given the total number of substitutions in the tree.
Even though these subtree tests are intuitively appealing, they suffer from the same problems discussed previously for the all-branch test \citep{siepel2006new}.
\par
The SPH marginal subtree test can be corrected and made more powerful by using the correct asymptotic distribution of $\mathcal{T}_{sub}$ as the null distribution; this asymptotic distribution is obtained using reasoning similar to that used for the asymptotic distribution of $\mathcal{T}_{all}$.
We propose our own conditional subtree test based on the following test statistic:
\begin{align}
\begin{split}
\mathcal{T}_{ratio} \equiv \mathcal{T}_{ratio}(\mathbf{D}_{1:L}) = \frac{\mathcal{T}_{sub}(\mathbf{D}_{1:L})}{\mathcal{T}_{all}(\mathbf{D}_{1:L})} &= \text{E}\Biggl(\sum_{i=1}^L H^{(i)}_{\Theta_b} {\Bigg |} \mathbf{D}_{1:L}\Biggr) {\Bigg /} \text{E}\Biggl(\sum_{i=1}^L H^{(i)}_\Theta {\Bigg |} \mathbf{D}_{1:L}\Biggr) \\
&= \Biggl[\sum_{i=1}^L \text{E}(H_{\Theta_b} | \mathbf{D}_i)\Biggr] {\Bigg /} \Biggl[\sum_{i=1}^L \text{E}(H_\Theta | \mathbf{D}_i)\Biggr],
\end{split}
\label{eq37}
\end{align}
where $H^{(i)}_{\Theta_b}$ and $H^{(i)}_\Theta$ are defined as previously.
This test statistic serves as a proxy for the ``observed'' proportion of substitutions in the subtree defined by $\Theta_b$ across the alignment $\mathbf{D}_{1:L}$.
Using the Central Limit Theorem and the Delta Method, we approximate the sampling distribution of $\mathcal{T}_{ratio}$ with a normal distribution whose moments depend on $\text{E}(H_{\Theta_b})$, $\text{E}(H_\Theta)$, $\text{Var}[\text{E}(H_{\Theta_b} | \mathbf{D})]$, $\text{Var}[\text{E}(H_\Theta | \mathbf{D})]$, and $\text{Cov}[\text{E}(H_{\Theta_b} | \mathbf{D}), \text{E}(H_\Theta | \mathbf{D})]$; we omit the exact forms of the mean and variance of this asymptotic distribution for brevity.
By the Laws of Total Variance and Covariance, we can express $\text{Var}[\text{E}(H_{\Theta_b} | \mathbf{D})]$, $\text{Var}[\text{E}(H_\Theta | \mathbf{D})]$, and $\text{Cov}[\text{E}(H_{\Theta_b} | \mathbf{D}), \text{E}(H_\Theta | \mathbf{D})]$ as follows:
\begin{align}
\text{Var}[\text{E}(H_{\Theta_b} | \mathbf{D})] &= \text{Var}(H_{\Theta_b}) - \text{E}[\text{Var}(H_{\Theta_b} | \mathbf{D})],
\label{eq38} \\
\text{Var}[\text{E}(H_\Theta | \mathbf{D})] &= \text{Var}(H_\Theta) - \text{E}[\text{Var}(H_\Theta | \mathbf{D})],
\label{eq39} \\
\text{Cov}[\text{E}(H_{\Theta_b} | \mathbf{D}), \text{E}(H_\Theta | \mathbf{D})] &= \text{Cov}(H_{\Theta_b}, H_\Theta) - \text{E}[\text{Cov}(H_{\Theta_b}, H_\Theta | \mathbf{D})].
\label{eq40}
\end{align}
The prior moments $\text{E}(H_{\Theta_b})$, $\text{E}(H_\Theta)$, $\text{Var}(H_{\Theta_b})$, $\text{Var}(H_\Theta)$, and $\text{Cov}(H_{\Theta_b}, H_\Theta)$ are efficiently computed using the post-order tree traversal algorithm outlined in the Appendix.
The quantities $\text{E}[\text{Var}(H_{\Theta_b} | \mathbf{D})]$, $\text{E}[\text{Var}(H_\Theta | \mathbf{D})]$, and $\text{E}[\text{Cov}(H_{\Theta_b}, H_\Theta | \mathbf{D})]$ are approximated using Monte Carlo replicates of $\mathbf{D}$ and our exact calculations of posterior mapping variances and covariances.
For both of our lineage-specific conservation tests, we estimate the global scale parameter $\rho$ by numerically maximizing the observed log-likelihood function.
In the next subsubsection, we present simulation results that demonstrate the utility of our modified SPH conservation tests.

\subsubsection{Simulation Experiments}

We evaluate the power and false positive rates of the original and modified SPH conservation tests using simulated alignments.
The neutral evolutionary model used by \citet{pollard2010detection} is also employed in all our simulation experiments.
This model was estimated using fourfold degenerate sites extracted from alignments of the 44 ENCODE regions \citep{encode2007identification} for 36 vertebrate species.
For the all-branch and subtree tests, we simulate replicate alignments by independently generating $L$ alignment columns according to $\psi(\rho)$ and $\psi(\rho, \lambda; \Theta_{primates})$, respectively, where $primates$ denotes the branch above the primates subtree in the neutral phylogeny.
We consider $L = 1, 2, 4, ..., 48, 50$ and $\rho = 0.1, 0.3, 0.5, 0.7, 0.9, 1$ in our all-branch simulations and $L = 1, 5, 10, ..., 45, 50$; $\rho = 0.1, 0.25, 0.4, ..., 1$; and $\lambda = 0.1, 0.25, 0.4, ..., 1$ in our subtree simulations.
We keep the alignment length $L$ relatively small because the primary application of SPH tests is scanning whole genomes in search of short ultra-conserved genetic elements.
For each simulation setting under $H_a$, we generate 1000 replicate datasets, compute the conservation $p$-values for each dataset using the original and modified SPH tests, and estimate the power by calculating the proportion of $p$-values less than the given significance level; we construct power curves by varying the significance threshold between 0 and 1.
False positive rates are similarly estimated for each simulation setting under $H_0$.
Note that these power curves should not be confused with receiver operating characteristic (ROC) curves; in our simulations, we plot the power (i.e.\ true positive rates) against the significance levels, whereas ROC curves plot the power against the false positive rates.
\begin{figure}[!ht]
\centering
\includegraphics[width=\textwidth, trim=0 10 0 5, clip]{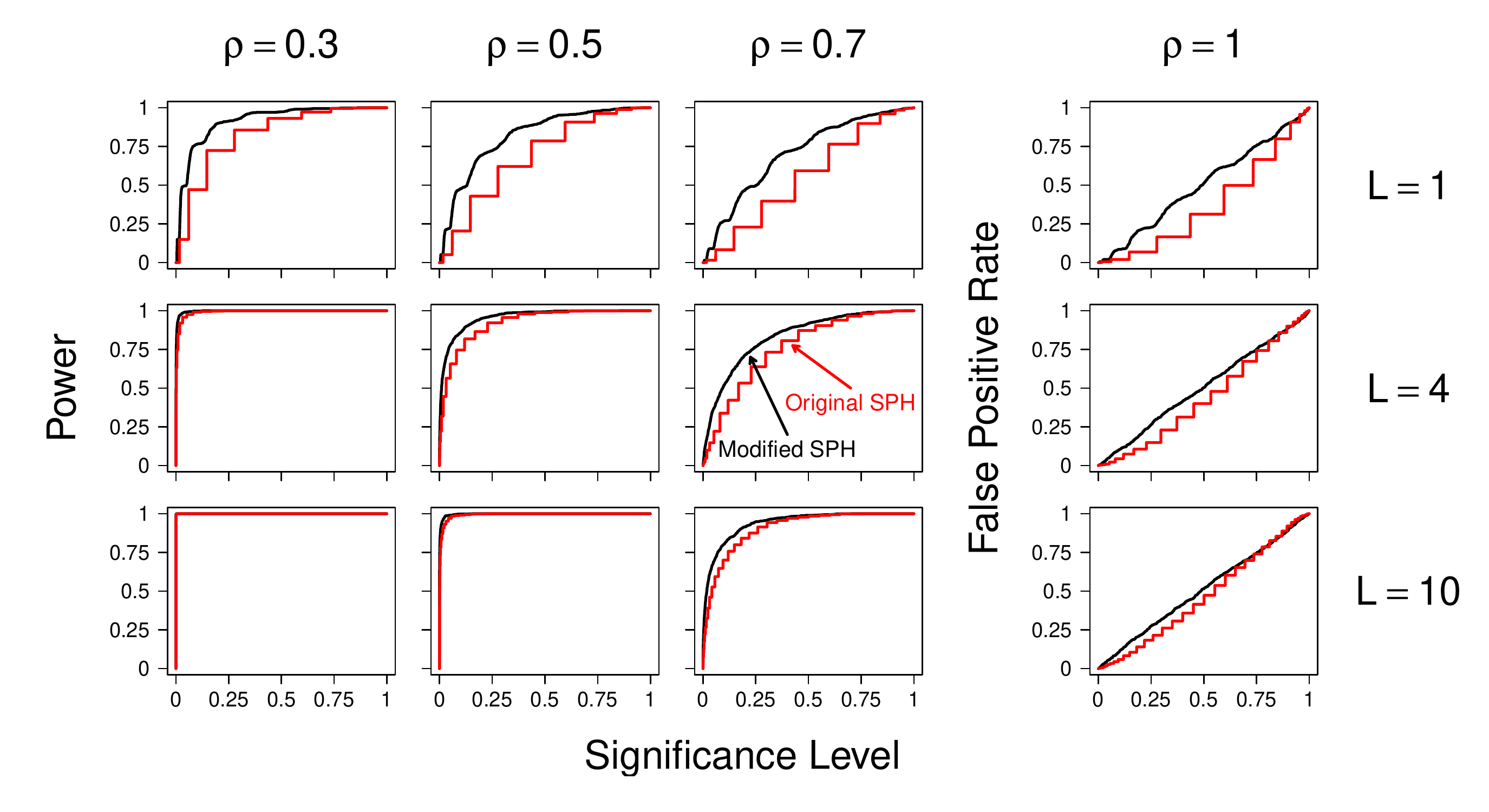}
\caption{Power and false positive rate plots from our all-branch simulation experiments.  The power and false positive rate curves for the original SPH all-branch test are shown in red, while the corresponding performance curves for the modified SPH all-branch test are displayed in black.  In this figure, we present performance plots for $L = 1, 4, 10$ and $\rho = 0.3, 0.5, 0.7, 1$.}
\label{allbranch}
\end{figure}
\par
In \autoref*{allbranch}, we display some of the power and false positive rate plots from our all-branch simulation experiments.
Specifically, we present performance plots for $L = 1, 4, 10$ and $\rho = 0.3, 0.5, 0.7, 1$.
These plots suggest that the modified SPH all-branch test is more powerful than the original SPH all-branch test.
The gap between the power curves for the two tests is negligible for large $L$ and small $\rho$ but increases as we examine shorter alignments with lower levels of conservation.
This latter result is surprising because the null distribution used in the modified SPH all-branch test is based on an asymptotic approximation.
The false positive rate plots seem to indicate that the $p$-values obtained from the modified all-branch test are approximately uniformly distributed under the null hypothesis, even for small $L$.
In addition, it is apparent that the original SPH all-branch $p$-values are conservative under $H_0$, confirming the results found in \citep{siepel2006new}.
\begin{figure}[!ht]
\begin{subfigure}{\textwidth}
\centering
\includegraphics[width=\textwidth, trim=0 10 0 10, clip]{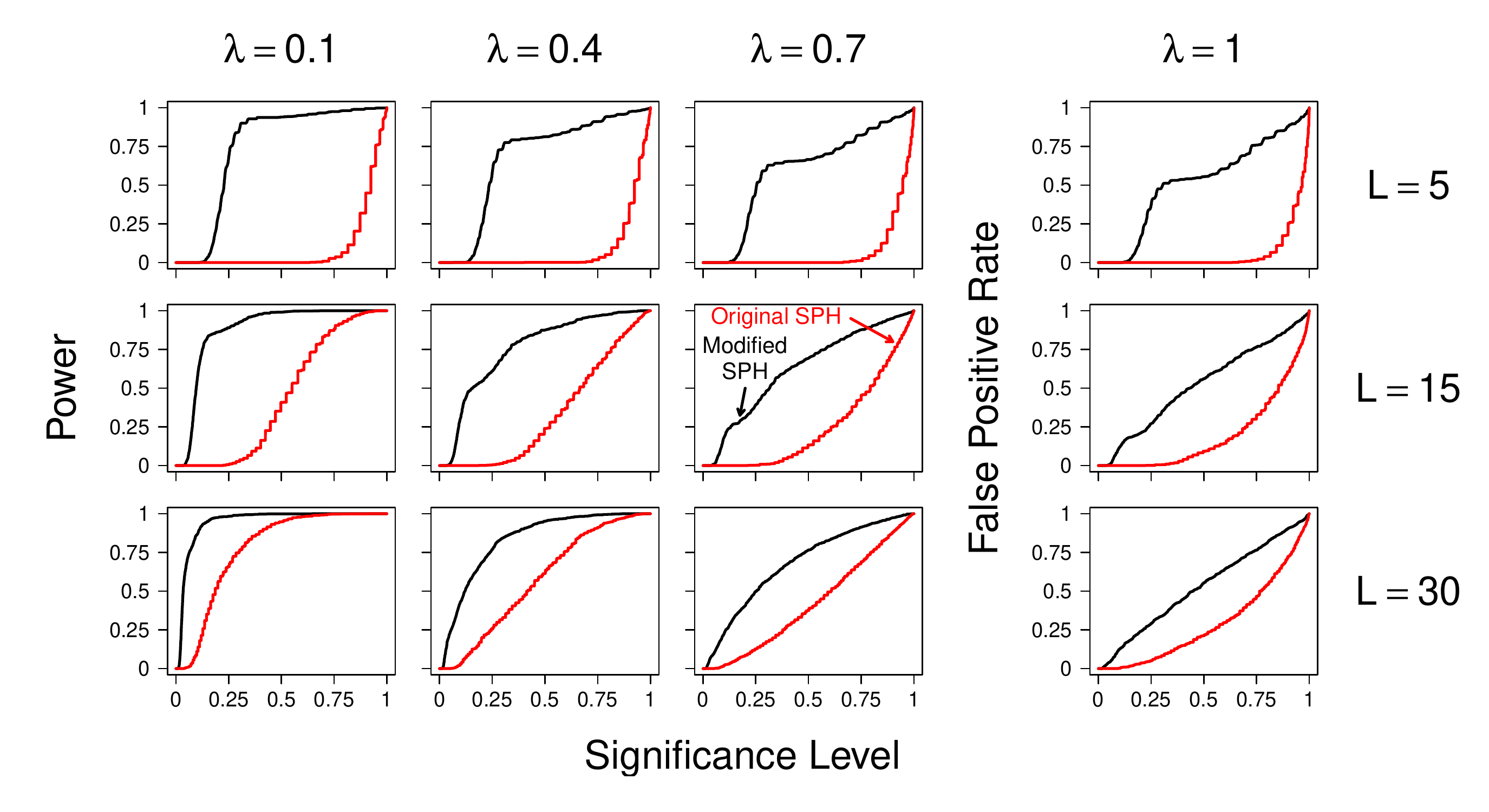}
\caption{Power and false positive rate curves for $\rho = 0.25$}
\label{subtreesmallrho}
\end{subfigure}
\newline
\vspace{5mm}
\newline
\begin{subfigure}{\textwidth}
\centering
\includegraphics[width=\textwidth, trim=0 10 0 0, clip]{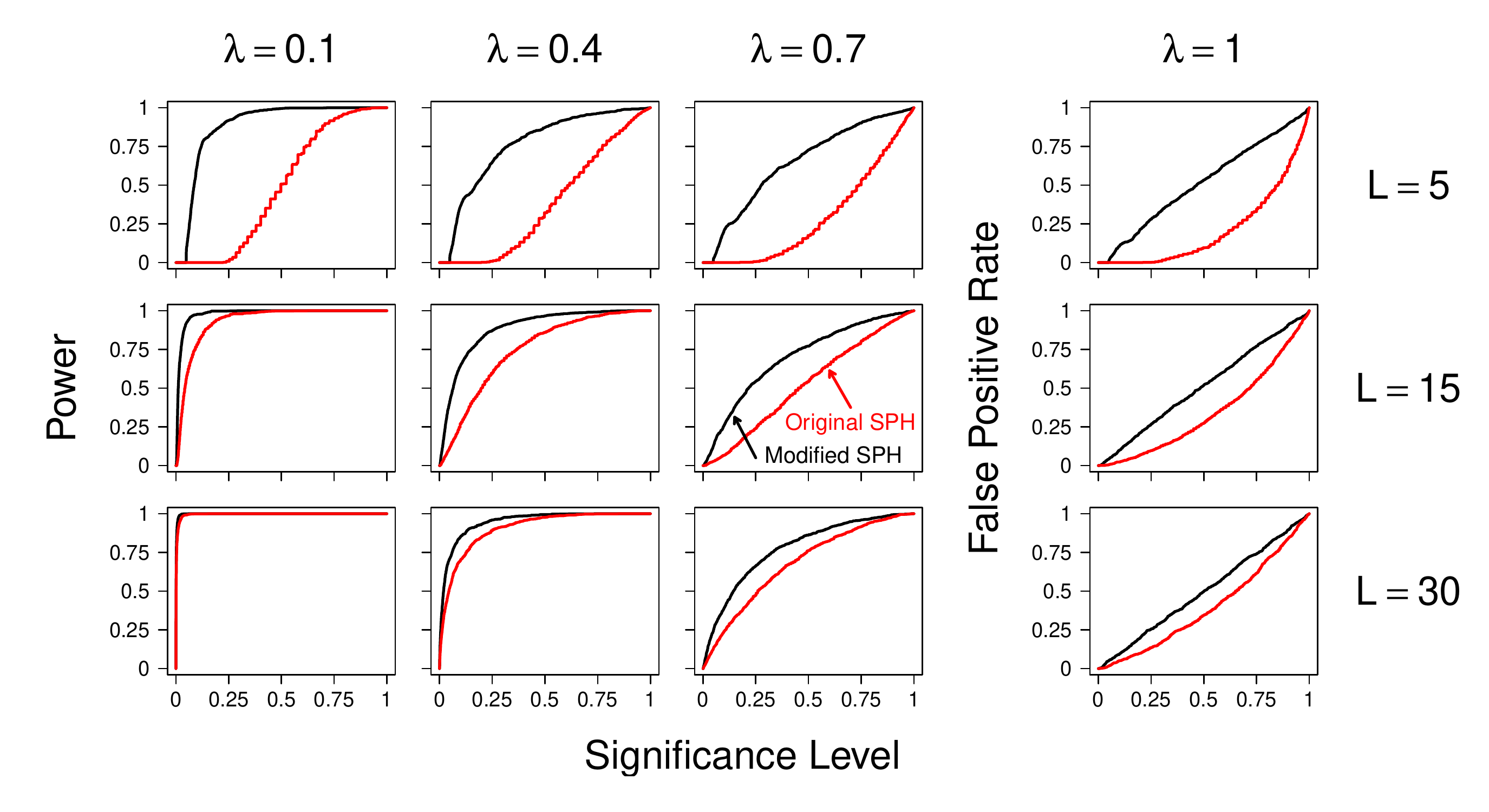}
\caption{Power and false positive rate curves for $\rho = 0.85$}
\label{subtreelargerho}
\end{subfigure}
\caption{Power and false positive rate plots from our subtree simulation experiments.  The power and false positive rate curves for the original SPH conditional subtree test are shown in red, while the corresponding performance curves for the modified SPH conditional subtree test are displayed in black.  In this figure, we present performance plots for $L = 5, 15, 30$; $\rho = 0.25, 0.85$; and $\lambda = 0.1, 0.4, 0.7, 1$.}
\label{subtree}
\end{figure}
\par
\autoref*{subtree} presents power and false positive rate plots from our subtree simulations.
We provide performance plots for $L = 5, 15, 30$; $\rho = 0.25, 0.85$; and $\lambda = 0.1, 0.4, 0.7, 1$.
The modified SPH subtree tests are more powerful than the original SPH subtree tests in all our simulation experiments.
Furthermore, we find that the power curves for the two modified subtree tests are nearly identical; the power curves for the two original subtree tests are similar as well.
We only display the power curves for the conditional subtree tests in \autoref*{subtree}; the full set of power curves is shown in \autoref*{subtreesupp} (see Appendix).
\par
Our subtree simulation experiments suggest that the presence of strong phylogeny-wide conservation (as measured by $\rho$) makes it more difficult to correctly identify lineage-specific conservation.
For $\rho$ close to 1, the separation between the power curves for the original and modified subtree tests is minimal when $L$ is large and $\lambda$ is small but widens as we analyze shorter elements with lower levels of primate-specific conservation.
However, for $\rho$ close to 0, the power curves differ quite substantially for all settings of $L$ and $\lambda$.
The power of the modified subtree tests is more robust to changes in $\rho$, while the power of the original subtree tests diminishes greatly as $\rho$ decreases.
Thus, it appears that the modified subtree tests can more accurately detect lineage-specific conservation in the presence of strong phylogeny-wide conservation.
\par
Our results show that testing for conservation in a subtree of interest is more difficult than testing for conservation across the entire phylogeny.
This is not too surprising because we have to account for the uncertainty associated with estimating $\rho$ in the subtree tests and the modified subtree test statistics achieve asymptotic normality under $H_0$ at a rate slower than is observed for the modified all-branch test statistic.
The latter can be seen by comparing the false positive rate plots in Figures \ref{allbranch} and \ref{subtree}; additionally, these plots indicate that the original SPH subtree $p$-values are more conservative than the original SPH all-branch $p$-values.

\section{Discussion}

In this paper, we present a post-order tree traversal algorithm that computes prior and posterior stochastic mapping variances with space and time complexity linear in the number of tips on the phylogeny; prior and posterior mapping covariances are efficiently calculated using a generalized version of this algorithm (see Appendix).
In many applications, including the ones presented in this paper, the posterior distribution of stochastic mapping summaries can be approximated by a normal distribution.
Since the normal distribution is fully specified by its mean vector and covariance matrix, our new stochastic mapping (co)variance computation together with an already available efficient way of computing the stochastic mapping mean enables access to the full posterior distribution of stochastic mapping summaries without resorting to costly simulations.
Our methodology builds upon the results of \citet{minin2008fast} and is inspired by the work of \citet{kenney2012hessian}, who devised a dynamic programming procedure for calculating second derivatives of phylogenetic likelihood functions.
\par
In fact, our algorithm for computing prior and posterior mapping moments can be adapted to provide a more straightforward description of the \citet{kenney2012hessian} algorithm.
If we replace the restricted factorial moments in the recursive equations of our algorithm with the appropriate derivatives of CTMC transition probabilities, then we would obtain an algorithm that computes the second derivatives of interest.
Even though this reformulation is equivalent to the approach of \citet{kenney2012hessian}, we believe our presentation is more streamlined and easier to follow.
We also point out that our algorithm and the work by \citet{kenney2012hessian} can be viewed as extensions of analogous calculations for hidden Markov models \citep{lystig2002, cappe2005}.
\par
The exact calculations of prior and posterior mapping moments allow us to construct more efficient posterior predictive rate variation tests and more accurate tests of genomic conservation.
From our analyses of the $\beta$-globin and influenza datasets, we find that the posterior dispersion index for substitution counts is a useful discrepancy measure for detecting observed rate variation across sites.
We motivate our use of the posterior dispersion index as a discrepancy measure by alluding to the prior dispersion index for substitution counts often being greater than 1 in the presence of rate variation among sites.
The observed and predicted distributions of the posterior dispersion index in \autoref*{obspreddist} do not have support outside the interval $(0,1)$, but this is not entirely inconsistent with the reasoning given above because a simple limiting argument shows that the posterior dispersion index for long sequence alignments is less than or equal to the prior dispersion index, even in the presence of across-site rate variation.
From our all-branch and subtree simulation experiments, we see that the modified SPH tests are better than the original SPH tests at correctly identifying phylogeny-wide and lineage-specific conservation in genomic sequence alignments.
Specifically, we observe that the differences in power between the original and modified conservation tests are greatest when we analyze short elements with low levels of conservation.
Similarly to the techniques found in \citep{kellis2003sequencing}, our modified tests of conservation could aid in the discovery of new transcriptional regulatory motifs in the human genome.
\par
The work presented here can be extended in several different directions.
One obvious extension is to generalize our post-order tree traversal algorithm to calculate higher-order moments of stochastic mapping summaries, such as the (co)skewness and (co)kurtosis.
These higher-order moments could then be used to construct more complex discrepancy measures for posterior predictive model diagnostics.
Similarly to the test of rate variation among sites, it is of interest to develop a posterior predictive approach to testing for rate variation among branches of the phylogeny.
Posterior predictive tests of this type could serve as useful diagnostic tools to assess the appropriateness of relaxed molecular clock models \citep{drummond2006relaxed}.
In addition, it would also be beneficial to establish a posterior predictive framework for testing the stationarity and homogeneity assumptions implicit in reversible substitution models.
These tests could be used to determine the suitability of nonreversible substitution models \citep{boussau2006efficient}.
Our dynamic programming algorithm can be easily altered to compute the posterior mapping variance of labeled dwelling times, which could be employed as a discrepancy measure in these posterior predictive tests.
Finally, another potential avenue for future research is to improve the two tests of genomic acceleration discussed by \citet{pollard2010detection}, where acceleration refers to evolution that is faster than expected.
Pursuing this line of research could lead to the detection of new ``human accelerated regions'' in our genome \citep{pollard2006forces, pollard2006rna}.
Based on the results in this paper and the promising directions for further study, we think stochastic mapping is and will continue to be essential to making reliable inferences about the latent evolutionary process on the phylogeny.

\section*{Acknowledgments}

We thank Melissa Hubisz for providing us access to the neutral evolutionary model used in \citep{pollard2010detection} and answering our questions about the R package \verb|rphast|.
VNM was supported in part by the National Institute of Health grant R01-AI107034.

\bibliographystyle{apa}
\bibliography{biblio.bib}

\newpage
\section*{Appendix}

\subsection*{Monte Carlo Summary Tables}

\begin{table}[!ht]
\begin{subtable}{\textwidth}
\centering
\begin{tabular}{cccccccccc}
\hline
& \multicolumn{4}{c}{Standard errors ($\beta$-globin)} & & \multicolumn{4}{c}{Standard errors (influenza)} \\
\cline{2-5} \cline{7-10}
& $L = 50$ & $L = 100$ & $L = 200$ & $L = 400$ & & $L = 50$ & $L = 100$ & $L = 200$ & $L = 400$ \\
\hline
$m = 100$ & 0.91 & 1.3 & 1.8 & 2.5 & & 0.094 & 0.13 & 0.19 & 0.26 \\
$m = 500$ & 0.41 & 0.57 & 0.82 & 1.2 & & 0.042 & 0.060 & 0.086 & 0.12 \\
$m = 1000$ & 0.29 & 0.41 & 0.58 & 0.82 & & 0.030 & 0.042 & 0.061 & 0.085 \\
$m = 10000$ & 0.092 & 0.13 & 0.18 & 0.26 & & 0.0097 & 0.013 & 0.019 & 0.027 \\
\hline
\end{tabular}
\caption{Monte Carlo standard errors for the $\beta$-globin and influenza datasets}
\label{mcerr}
\end{subtable}
\newline
\vspace{2.5mm}
\newline
\begin{subtable}{\textwidth}
\centering
\begin{tabular}{ccccc}
\hline
& $L = 50$ & $L = 100$ & $L = 200$ & $L = 400$ \\
\hline
$\beta$-globin & 22 & 41 & 84 & 170 \\
influenza & 0.34 & 0.54 & 1.3 & 2.3 \\
\hline
\end{tabular}
\caption{Exact calculations of $T_{var}$ for the $\beta$-globin and influenza datasets}
\label{exactvar}
\end{subtable}
\caption{Monte Carlo summary tables for the $\beta$-globin and influenza datasets.  (a) Monte Carlo standard errors associated with simulation-based $T_{var}$ estimates.  We compute these standard errors on randomly subsampled alignments of length $L$ using $m$ Monte Carlo replicates per site.  Each table entry represents an averaged Monte Carlo standard error, where the averaging is done over 200 randomly subsampled posterior $\boldsymbol{\theta}$'s.  (b) Exact computations of $T_{var}$.  Each table entry denotes an average over exact values of $T_{var}$, where the averaging is done over the same 200 posterior samples of $\boldsymbol{\theta}$ mentioned above.  All table entries in (a) and (b) are rounded to two significant digits.}
\label{mctablesupp}
\end{table}
\vspace{-3.75mm}

\subsection*{Prior and Posterior Mapping Covariance Computation}

In this subsection, we describe how to efficiently compute prior and posterior mapping covariances.
We generalize the post-order tree traversal algorithm discussed in the main part of the paper and present the necessary formulas for calculating these covariances.
Much of the intuition provided for our original tree traversal algorithm carries over to this generalized procedure.
\par
Let $\text{Cov}(H_{\Omega_1}, H_{\Omega_2})$ and $\text{Cov}(H_{\Omega_1}, H_{\Omega_2} | \mathbf{D})$ denote the prior and posterior mapping covariances, respectively, where $\Omega_1, \Omega_2 \subseteq \Theta$ are predefined sets of branch indices.
We consider first the calculation of the posterior mapping covariance $\text{Cov}(H_{\Omega_1}, H_{\Omega_2} | \mathbf{D})$.
The vectors $\mathbf{F}_u$ and $\mathbf{S}_b$ are defined and computed as in our original tree traversal procedure for all nodes $u \in \{ 1, ..., n-1, n, ..., 2n-1 \}$ and branches $b \in \Theta$.
We introduce the $m$-long vectors $\mathbf{V}^{(\Omega_1,[1])}_b$, $\mathbf{V}^{(\Omega_1,[2])}_b$, $\mathbf{V}^{(\Omega_2,[1])}_b$, $\mathbf{V}^{(\Omega_2,[2])}_b$, $\mathbf{V}^{(\Omega_1 \cap \Omega_2,[1])}_b$, $\mathbf{V}^{(\Omega_1 \cap \Omega_2,[2])}_b$, $\mathbf{W}^{(\Omega_1)}_b$, $\mathbf{W}^{(\Omega_2)}_b$, and $\mathbf{W}^{(\Omega_1,\Omega_2)}_b$ for all $b \in \Theta$.
The $i$th entries in $\mathbf{V}^{(\Omega_1,[1])}_b$ and $\mathbf{V}^{(\Omega_1,[2])}_b$ are mathematically defined as:
\begin{gather}
\sum_{b^\dagger} \sum_{\mathbf{i}_b} e^{[1]}_{i^*_{p(b^\dagger)} i^*_{c(b^\dagger)}}(h, t_{b^\dagger}) \prod_{b^* \in \Theta_b \setminus \{ b^\dagger \}} p_{i^*_{p(b^*)} i^*_{c(b^*)}}(t_{b^*}),
\label{eq41} \\[10pt]
\sum_{b^\dagger} \sum_{\mathbf{i}_b} e^{[2]}_{i^*_{p(b^\dagger)} i^*_{c(b^\dagger)}}(h, t_{b^\dagger}) \prod_{b^* \in \Theta_b \setminus \{ b^\dagger \}} p_{i^*_{p(b^*)} i^*_{c(b^*)}}(t_{b^*}),
\label{eq42}
\end{gather}
respectively, where the state of parent node $p(b)$ is $i$ and $b^\dagger \in \Omega_{1,b}$ for $\Omega_{1,b} = \Omega_1 \cap \Theta_b$.
The entries in $\mathbf{V}^{(\Omega_2,[1])}_b$ and $\mathbf{V}^{(\Omega_2,[2])}_b$ and $\mathbf{V}^{(\Omega_1 \cap \Omega_2,[1])}_b$ and $\mathbf{V}^{(\Omega_1 \cap \Omega_2,[2])}_b$ are defined analogously by replacing $\Omega_1$ with $\Omega_2$ and $\Omega_1 \cap \Omega_2$, respectively, in the above definitions.
The $i$th element of the vector $\mathbf{W}^{(\Omega_1)}_b$ is equal to:
\begin{equation}
\sum_{b^\dagger \neq b^{\dagger \dagger}} \sum_{\mathbf{i}_b} e^{[1]}_{i^*_{p(b^\dagger)} i^*_{c(b^\dagger)}}(h, t_{b^\dagger}) e^{[1]}_{i^*_{p(b^{\dagger \dagger})} i^*_{c(b^{\dagger \dagger})}}(h, t_{b^{\dagger \dagger}}) \prod_{b^* \in \Theta_b \setminus \{ b^\dagger,b^{\dagger \dagger} \}} p_{i^*_{p(b^*)} i^*_{c(b^*)}}(t_{b^*}),
\label{eq43}
\end{equation}
where the state of parent node $p(b)$ is $i$ and $b^\dagger,b^{\dagger \dagger} \in \Omega_{1,b}$ for $\Omega_{1,b}$ defined as above.
The elements in $\mathbf{W}^{(\Omega_2)}_b$ and $\mathbf{W}^{(\Omega_1,\Omega_2)}_b$ are similarly defined, except that $b^\dagger,b^{\dagger \dagger} \in \Omega_{2,b}$ and $b^\dagger \in \Omega_{1,b}, b^{\dagger \dagger} \in \Omega_{2,b}$, respectively, for the same $\Omega_{1,b}$ and $\Omega_{2,b} = \Omega_2 \cap \Theta_b$.
\par
For all terminal branches $b \in \mathcal{E}$, we define:
\begin{gather}
\mathbf{V}^{(\Omega_1,[1])}_b = \mathbf{e}^{[1]}(h, t_b) \mathbf{F}_{c(b)} \mathbbm{1}_{\{ b \in \Omega_1 \}},
\label{eq44} \\[10pt]
\mathbf{V}^{(\Omega_1,[2])}_b = \mathbf{e}^{[2]}(h, t_b) \mathbf{F}_{c(b)} \mathbbm{1}_{\{ b \in \Omega_1 \}},
\label{eq45} \\[10pt]
\mathbf{V}^{(\Omega_2,[1])}_b = \mathbf{e}^{[1]}(h, t_b) \mathbf{F}_{c(b)} \mathbbm{1}_{\{ b \in \Omega_2 \}},
\label{eq46} \\[10pt]
\mathbf{V}^{(\Omega_2,[2])}_b = \mathbf{e}^{[2]}(h, t_b) \mathbf{F}_{c(b)} \mathbbm{1}_{\{ b \in \Omega_2 \}},
\label{eq47} \\[10pt]
\mathbf{V}^{(\Omega_1 \cap \Omega_2,[1])}_b = \mathbf{e}^{[1]}(h, t_b) \mathbf{F}_{c(b)} \mathbbm{1}_{\{ b \in \Omega_1 \cap \Omega_2 \}},
\label{eq48} \\[10pt]
\mathbf{V}^{(\Omega_1 \cap \Omega_2,[2])}_b = \mathbf{e}^{[2]}(h, t_b) \mathbf{F}_{c(b)} \mathbbm{1}_{\{ b \in \Omega_1 \cap \Omega_2 \}}.
\label{eq49}
\end{gather}
All entries in $\mathbf{W}^{(\Omega_1)}_b$, $\mathbf{W}^{(\Omega_2)}_b$, and $\mathbf{W}^{(\Omega_1,\Omega_2)}_b$ for $b \in \mathcal{E}$ are set to 0.
The recursive formulas for calculating these vectors at internal branches $b \in \mathcal{I}$ are:
\begin{gather}
\mathbf{V}^{(\Omega_1,[1])}_b = \mathbf{e}^{[1]}(h, t_b) \mathbf{F}_{c(b)} \mathbbm{1}_{\{ b \in \Omega_1 \}} + \mathbf{P}(t_b) \Bigl(\mathbf{V}^{(\Omega_1,[1])}_{b_1} \circ \mathbf{S}_{b_2} + \mathbf{V}^{(\Omega_1,[1])}_{b_2} \circ \mathbf{S}_{b_1}\Bigr),
\label{eq50} \\[10pt]
\mathbf{V}^{(\Omega_1,[2])}_b = \mathbf{e}^{[2]}(h, t_b) \mathbf{F}_{c(b)} \mathbbm{1}_{\{ b \in \Omega_1 \}} + \mathbf{P}(t_b) \Bigl(\mathbf{V}^{(\Omega_1,[2])}_{b_1} \circ \mathbf{S}_{b_2} + \mathbf{V}^{(\Omega_1,[2])}_{b_2} \circ \mathbf{S}_{b_1}\Bigr),
\label{eq51} \\[10pt]
\mathbf{V}^{(\Omega_2,[1])}_b = \mathbf{e}^{[1]}(h, t_b) \mathbf{F}_{c(b)} \mathbbm{1}_{\{ b \in \Omega_2 \}} + \mathbf{P}(t_b) \Bigl(\mathbf{V}^{(\Omega_2,[1])}_{b_1} \circ \mathbf{S}_{b_2} + \mathbf{V}^{(\Omega_2,[1])}_{b_2} \circ \mathbf{S}_{b_1}\Bigr),
\label{eq52} \\[10pt]
\mathbf{V}^{(\Omega_2,[2])}_b = \mathbf{e}^{[2]}(h, t_b) \mathbf{F}_{c(b)} \mathbbm{1}_{\{ b \in \Omega_2 \}} + \mathbf{P}(t_b) \Bigl(\mathbf{V}^{(\Omega_2,[2])}_{b_1} \circ \mathbf{S}_{b_2} + \mathbf{V}^{(\Omega_2,[2])}_{b_2} \circ \mathbf{S}_{b_1}\Bigr),
\label{eq53} \\[10pt]
\mathbf{V}^{(\Omega_1 \cap \Omega_2,[1])}_b = \mathbf{e}^{[1]}(h, t_b) \mathbf{F}_{c(b)} \mathbbm{1}_{\{ b \in \Omega_1 \cap \Omega_2 \}} + \mathbf{P}(t_b) \Bigl(\mathbf{V}^{(\Omega_1 \cap \Omega_2,[1])}_{b_1} \circ \mathbf{S}_{b_2} + \mathbf{V}^{(\Omega_1 \cap \Omega_2,[1])}_{b_2} \circ \mathbf{S}_{b_1}\Bigr),
\label{eq54} \\[10pt]
\mathbf{V}^{(\Omega_1 \cap \Omega_2,[2])}_b = \mathbf{e}^{[2]}(h, t_b) \mathbf{F}_{c(b)} \mathbbm{1}_{\{ b \in \Omega_1 \cap \Omega_2 \}} + \mathbf{P}(t_b) \Bigl(\mathbf{V}^{(\Omega_1 \cap \Omega_2,[2])}_{b_1} \circ \mathbf{S}_{b_2} + \mathbf{V}^{(\Omega_1 \cap \Omega_2,[2])}_{b_2} \circ \mathbf{S}_{b_1}\Bigr),
\label{eq55} \\[10pt]
\begin{split}
\mathbf{W}^{(\Omega_1)}_b = \ &2 \times \mathbf{e}^{[1]}(h, t_b) \Bigl(\mathbf{V}^{(\Omega_1,[1])}_{b_1} \circ \mathbf{S}_{b_2} + \mathbf{V}^{(\Omega_1,[1])}_{b_2} \circ \mathbf{S}_{b_1}\Bigr) \mathbbm{1}_{\{ b \in \Omega_1 \}} \\
&+ \mathbf{P}(t_b) \Bigl(2 \times \mathbf{V}^{(\Omega_1,[1])}_{b_1} \circ \mathbf{V}^{(\Omega_1,[1])}_{b_2} + \mathbf{W}^{(\Omega_1)}_{b_1} \circ \mathbf{S}_{b_2} + \mathbf{W}^{(\Omega_1)}_{b_2} \circ \mathbf{S}_{b_1}\Bigr),
\end{split}
\label{eq56} \\[10pt]
\begin{split}
\mathbf{W}^{(\Omega_2)}_b = \ &2 \times \mathbf{e}^{[1]}(h, t_b) \Bigl(\mathbf{V}^{(\Omega_2,[1])}_{b_1} \circ \mathbf{S}_{b_2} + \mathbf{V}^{(\Omega_2,[1])}_{b_2} \circ \mathbf{S}_{b_1}\Bigr) \mathbbm{1}_{\{ b \in \Omega_2 \}} \\
&+ \mathbf{P}(t_b) \Bigl(2 \times \mathbf{V}^{(\Omega_2,[1])}_{b_1} \circ \mathbf{V}^{(\Omega_2,[1])}_{b_2} + \mathbf{W}^{(\Omega_2)}_{b_1} \circ \mathbf{S}_{b_2} + \mathbf{W}^{(\Omega_2)}_{b_2} \circ \mathbf{S}_{b_1}\Bigr),
\end{split}
\label{eq57} \\[10pt]
\begin{split}
\mathbf{W}^{(\Omega_1,\Omega_2)}_b = \ &\mathbf{e}^{[1]}(h, t_b) \Bigl(\mathbf{V}^{(\Omega_2,[1])}_{b_1} \circ \mathbf{S}_{b_2} + \mathbf{V}^{(\Omega_2,[1])}_{b_2} \circ \mathbf{S}_{b_1}\Bigr) \mathbbm{1}_{\{ b \in \Omega_1 \}} \\
&+ \mathbf{e}^{[1]}(h, t_b) \Bigl(\mathbf{V}^{(\Omega_1,[1])}_{b_1} \circ \mathbf{S}_{b_2} + \mathbf{V}^{(\Omega_1,[1])}_{b_2} \circ \mathbf{S}_{b_1}\Bigr) \mathbbm{1}_{\{ b \in \Omega_2 \}} \\
& \! \begin{alignedat}{2}
&+ \mathbf{P}(t_b) \Bigl(&&\mathbf{V}^{(\Omega_1,[1])}_{b_1} \circ \mathbf{V}^{(\Omega_2,[1])}_{b_2} + \mathbf{V}^{(\Omega_1,[1])}_{b_2} \circ \mathbf{V}^{(\Omega_2,[1])}_{b_1}\Bigr. \\
&&&+ \Bigl.\mathbf{W}^{(\Omega_1,\Omega_2)}_{b_1} \circ \mathbf{S}_{b_2} + \mathbf{W}^{(\Omega_1,\Omega_2)}_{b_2} \circ \mathbf{S}_{b_1}\Bigr),
\end{alignedat}
\end{split}
\label{eq58}
\end{gather}
where $b_1$ and $b_2$ represent the two branches that are ``below'' branch $b$.
This generalized tree traversal algorithm terminates after computing $\mathbf{F}_u$, $\mathbf{S}_b$, $\mathbf{V}^{(\Omega_1,[1])}_b$, $\mathbf{V}^{(\Omega_1,[2])}_b$, $\mathbf{V}^{(\Omega_2,[1])}_b$, $\mathbf{V}^{(\Omega_2,[2])}_b$, $\mathbf{V}^{(\Omega_1 \cap \Omega_2,[1])}_b$, $\mathbf{V}^{(\Omega_1 \cap \Omega_2,[2])}_b$, $\mathbf{W}^{(\Omega_1)}_b$, $\mathbf{W}^{(\Omega_2)}_b$, and $\mathbf{W}^{(\Omega_1,\Omega_2)}_b$ for $u = root$ and $b \in \{ root_1,root_2 \}$, where $root$ denotes the root node label and $root_1$ and $root_2$ represent the two branches connecting the root node to its children.
The restricted mapping moments of interest are calculated as follows:
\begin{gather}
\text{E}(H_{\Omega_1} \mathbbm{1}_{\mathbf{D}}) = \boldsymbol{\pi}^T \Bigl(\mathbf{V}^{(\Omega_1,[1])}_{root_1} \circ \mathbf{S}_{root_2} + \mathbf{V}^{(\Omega_1,[1])}_{root_2} \circ \mathbf{S}_{root_1}\Bigr),
\label{eq59} \\[10pt]
\text{E}(H_{\Omega_2} \mathbbm{1}_{\mathbf{D}}) = \boldsymbol{\pi}^T \Bigl(\mathbf{V}^{(\Omega_2,[1])}_{root_1} \circ \mathbf{S}_{root_2} + \mathbf{V}^{(\Omega_2,[1])}_{root_2} \circ \mathbf{S}_{root_1}\Bigr),
\label{eq60} \\[10pt]
\begin{split}
\text{E}(H_{\Omega_1}^2 \mathbbm{1}_{\mathbf{D}}) = \boldsymbol{\pi}^T \Bigl[&2 \times \mathbf{V}^{(\Omega_1,[1])}_{root_1} \circ \mathbf{V}^{(\Omega_1,[1])}_{root_2} + \mathbf{W}^{(\Omega_1)}_{root_1} \circ \mathbf{S}_{root_2} + \mathbf{W}^{(\Omega_1)}_{root_2} \circ \mathbf{S}_{root_1}\Bigr. \\
&+ \Bigl.\bigl(\mathbf{V}^{(\Omega_1,[1])}_{root_1} + \mathbf{V}^{(\Omega_1,[2])}_{root_1}\bigr) \circ \mathbf{S}_{root_2} + \bigl(\mathbf{V}^{(\Omega_1,[1])}_{root_2} + \mathbf{V}^{(\Omega_1,[2])}_{root_2}\bigr) \circ \mathbf{S}_{root_1}\Bigr],
\end{split}
\label{eq61} \\[10pt]
\begin{split}
\text{E}(H_{\Omega_2}^2 \mathbbm{1}_{\mathbf{D}}) = \boldsymbol{\pi}^T \Bigl[&2 \times \mathbf{V}^{(\Omega_2,[1])}_{root_1} \circ \mathbf{V}^{(\Omega_2,[1])}_{root_2} + \mathbf{W}^{(\Omega_2)}_{root_1} \circ \mathbf{S}_{root_2} + \mathbf{W}^{(\Omega_2)}_{root_2} \circ \mathbf{S}_{root_1}\Bigr. \\
&+ \Bigl.\bigl(\mathbf{V}^{(\Omega_2,[1])}_{root_1} + \mathbf{V}^{(\Omega_2,[2])}_{root_1}\bigr) \circ \mathbf{S}_{root_2} + \bigl(\mathbf{V}^{(\Omega_2,[1])}_{root_2} + \mathbf{V}^{(\Omega_2,[2])}_{root_2}\bigr) \circ \mathbf{S}_{root_1}\Bigr],
\end{split}
\label{eq62} \\[10pt]
\begin{split}
\text{E}(H_{\Omega_1} H_{\Omega_2} \mathbbm{1}_{\mathbf{D}}) = \boldsymbol{\pi}^T \Bigl[&\mathbf{V}^{(\Omega_1,[1])}_{root_1} \circ \mathbf{V}^{(\Omega_2,[1])}_{root_2} + \mathbf{V}^{(\Omega_1,[1])}_{root_2} \circ \mathbf{V}^{(\Omega_2,[1])}_{root_1}\Bigr. \\
&+ \mathbf{W}^{(\Omega_1,\Omega_2)}_{root_1} \circ \mathbf{S}_{root_2} + \mathbf{W}^{(\Omega_1,\Omega_2)}_{root_2} \circ \mathbf{S}_{root_1} \\
&+ \bigl(\mathbf{V}^{(\Omega_1 \cap \Omega_2,[1])}_{root_1} + \mathbf{V}^{(\Omega_1 \cap \Omega_2,[2])}_{root_1}\bigr) \circ \mathbf{S}_{root_2} \\
&+ \Bigl.\bigl(\mathbf{V}^{(\Omega_1 \cap \Omega_2,[1])}_{root_2} + \mathbf{V}^{(\Omega_1 \cap \Omega_2,[2])}_{root_2}\bigr) \circ \mathbf{S}_{root_1}\Bigr].
\end{split}
\label{eq63}
\end{gather}
We also know that $\text{P}(\mathbf{D}) = \boldsymbol{\pi}^T \mathbf{F}_{root}$ \citep{felsenstein1981evolutionary}.
Our efficient computations of the above restricted mapping moments allow us to calculate the associated posterior mapping moments using the following equations:
\begin{gather}
\text{E}(H_{\Omega_1} | \mathbf{D}) = \frac{\text{E}(H_{\Omega_1} \mathbbm{1}_{\mathbf{D}})}{\text{P}(\mathbf{D})},
\label{eq64} \\[10pt]
\text{E}(H_{\Omega_2} | \mathbf{D}) = \frac{\text{E}(H_{\Omega_2} \mathbbm{1}_{\mathbf{D}})}{\text{P}(\mathbf{D})},
\label{eq65} \\[10pt]
\text{Var}(H_{\Omega_1} | \mathbf{D}) = \text{E}(H_{\Omega_1}^2 | \mathbf{D}) - \text{E}(H_{\Omega_1} | \mathbf{D})^2 = \frac{\text{E}(H_{\Omega_1}^2 \mathbbm{1}_{\mathbf{D}})}{\text{P}(\mathbf{D})} - \biggl[\frac{\text{E}(H_{\Omega_1} \mathbbm{1}_{\mathbf{D}})}{\text{P}(\mathbf{D})}\biggr]^2,
\label{eq66} \\[10pt]
\text{Var}(H_{\Omega_2} | \mathbf{D}) = \text{E}(H_{\Omega_2}^2 | \mathbf{D}) - \text{E}(H_{\Omega_2} | \mathbf{D})^2 = \frac{\text{E}(H_{\Omega_2}^2 \mathbbm{1}_{\mathbf{D}})}{\text{P}(\mathbf{D})} - \biggl[\frac{\text{E}(H_{\Omega_2} \mathbbm{1}_{\mathbf{D}})}{\text{P}(\mathbf{D})}\biggr]^2,
\label{eq67} \\[10pt]
\begin{split}
\text{Cov}(H_{\Omega_1}, H_{\Omega_2} | \mathbf{D}) &= \text{E}(H_{\Omega_1} H_{\Omega_2} | \mathbf{D}) - \text{E}(H_{\Omega_1} | \mathbf{D}) \text{E}(H_{\Omega_2} | \mathbf{D}) \\
&= \frac{\text{E}(H_{\Omega_1} H_{\Omega_2} \mathbbm{1}_{\mathbf{D}})}{\text{P}(\mathbf{D})} - \biggl[\frac{\text{E}(H_{\Omega_1} \mathbbm{1}_{\mathbf{D}})}{\text{P}(\mathbf{D})}\biggr] \biggl[\frac{\text{E}(H_{\Omega_2} \mathbbm{1}_{\mathbf{D}})}{\text{P}(\mathbf{D})}\biggr].
\end{split}
\label{eq68}
\end{gather}
\par
The prior mapping moments can be obtained by modifying the tree traversal procedure discussed in this subsection.
The only changes that need to be made are to the initializations of $\mathbf{F}_u$ at all terminal nodes $u$.
If we set $F_{ui} = 1$ for all terminal nodes $u \in \{ n, ..., 2n-1 \}$ and $i = 1, ..., m$, then our algorithm will be able to compute the prior moments of interest.
In this case, equations \eqref{eq64}-\eqref{eq68} are used to calculate the prior mapping moments.
Thus, the prior and posterior mapping covariances are computed in a similar fashion, but according to different initializations of the $\mathbf{F}_u$ vectors at the tips of the phylogeny.

\newpage
\subsection*{SPH Subtree Simulation Plots}

\begin{figure}[!ht]
\begin{subfigure}{\textwidth}
\centering
\includegraphics[width=\textwidth, trim=0 10 0 10, clip]{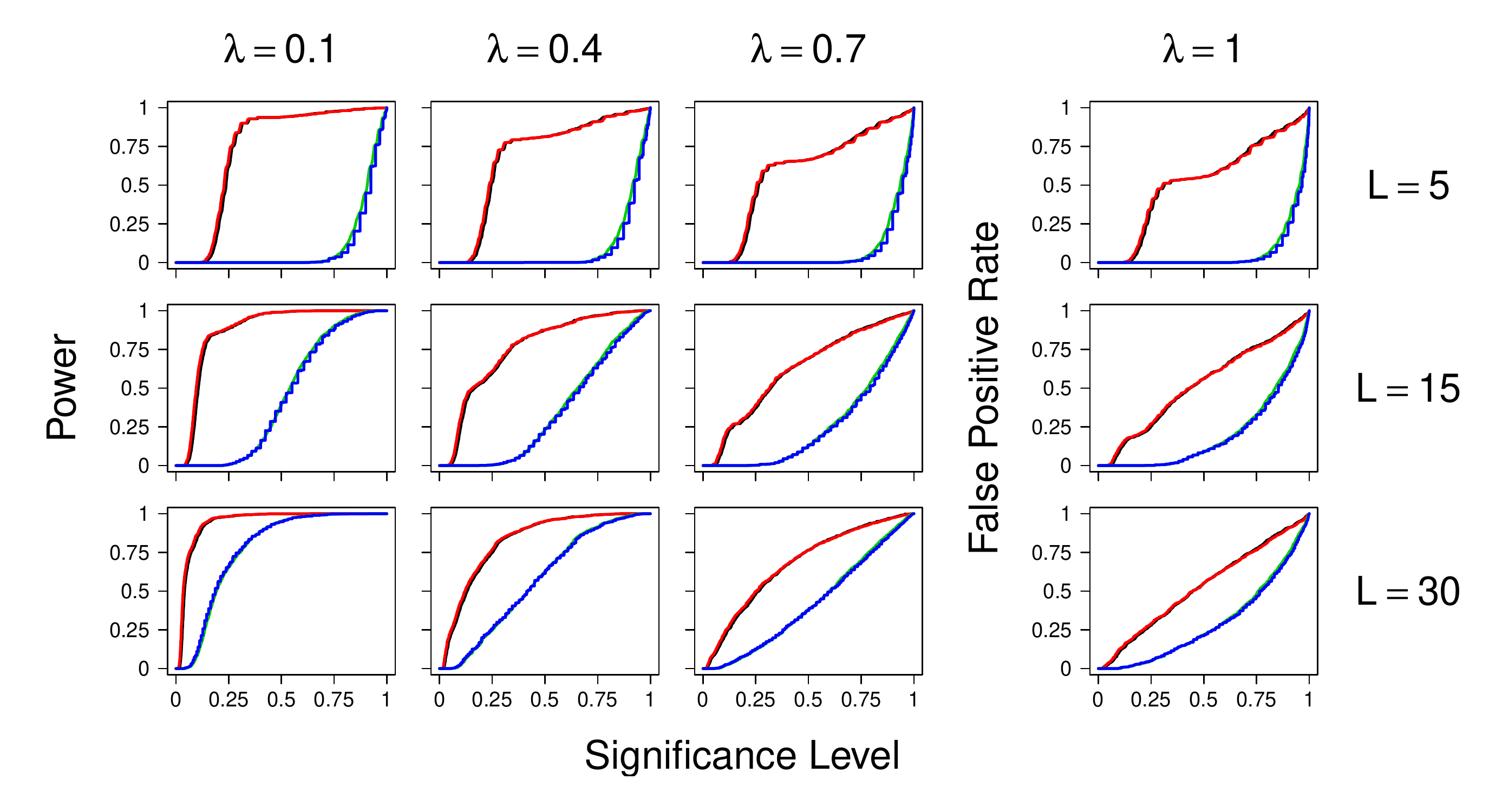}
\caption{Power and false positive rate curves for $\rho = 0.25$}
\label{subtreesmallrhosupp}
\end{subfigure}
\newline
\vspace{5mm}
\newline
\begin{subfigure}{\textwidth}
\centering
\includegraphics[width=\textwidth, trim=0 10 0 0, clip]{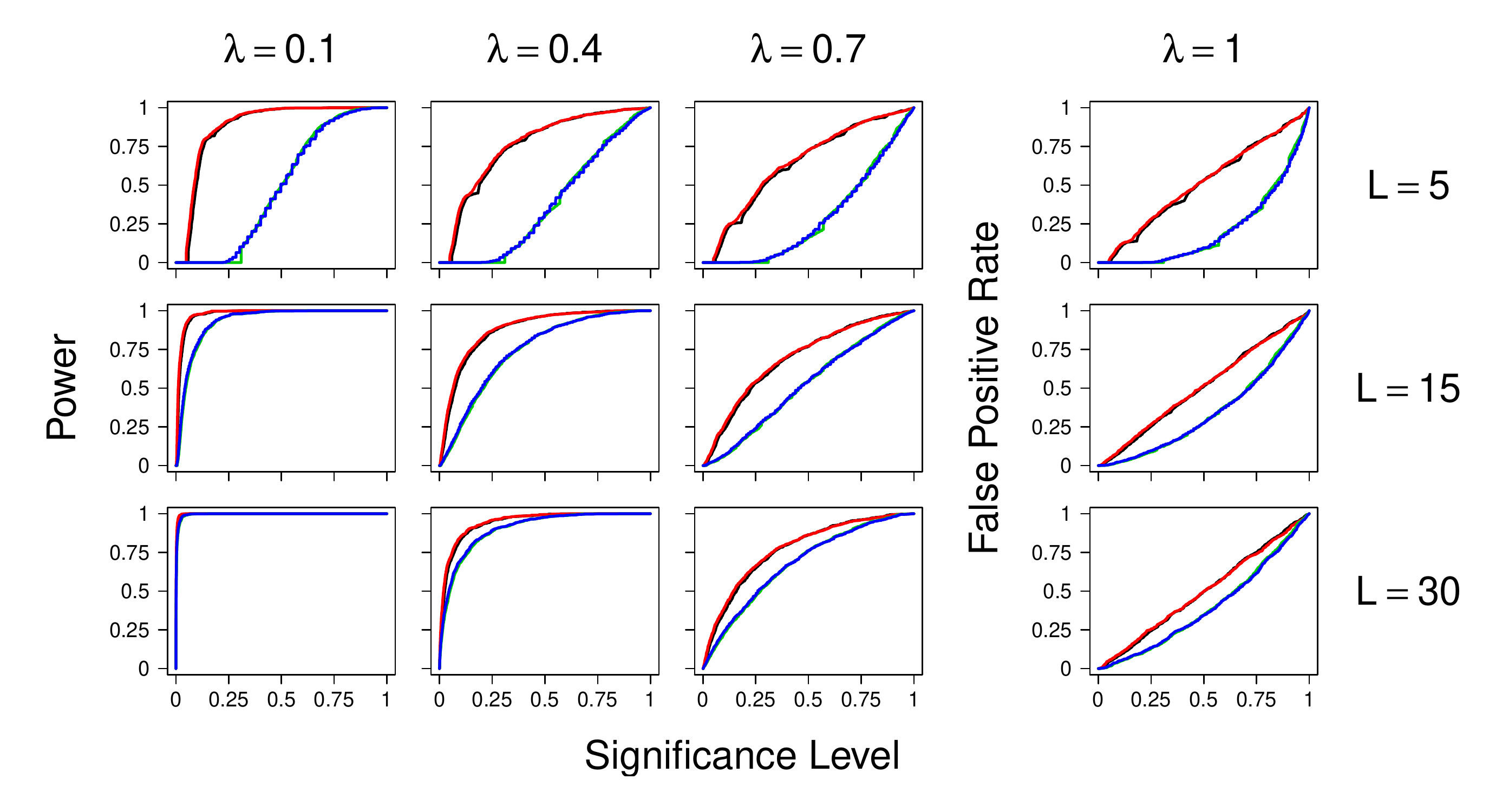}
\caption{Power and false positive rate curves for $\rho = 0.85$}
\label{subtreelargerhosupp}
\end{subfigure}
\caption{Power and false positive rate plots from our subtree simulation experiments.  The power and false positive rate curves for the original SPH marginal (conditional) subtree test are shown in green (blue), while the corresponding performance curves for the modified SPH marginal (conditional) subtree test are displayed in black (red).  In this figure, we present performance plots for $L = 5, 15, 30$; $\rho = 0.25, 0.85$; and $\lambda = 0.1, 0.4, 0.7, 1$.}
\label{subtreesupp}
\end{figure}

\end{document}